\documentclass{elsart-1}

\usepackage{graphicx}

\usepackage{amssymb,amsmath}

\usepackage{color}

\begin{document}

\begin{frontmatter}



\title{Finsler geometry modeling and Monte Carlo study of $3D$ liquid crystal elastomer}

\author{Keita Osari and Hiroshi Koibuchi}
\ead{koibuchi@mech.ibaraki-ct.ac.jp}

\address{Department of Mechanical and Systems Engineering, Ibaraki National College of Technology, 
Nakane 866, Hitachinaka,  Ibaraki 312-8508, Japan}

\begin{abstract}
We study a three-dimensional ($3D$) liquid crystal elastomer (LCE) in
the context of Finsler geometry (FG) modeling, where FG is a
mathematical framework for describing anisotropic phenomena. The LCE
is a $3D$ rubbery object and has remarkable properties, such as the
so-called soft elasticity and elongation, the mechanisms of which are
unknown at present.  To understand these anisotropic phenomena, we
introduce a variable $\sigma$, which represents the directional degrees
of freedom of a liquid crystal (LC) molecule. This variable $\sigma$
is used to define the Finsler metric for the interaction between the
LC molecules and bulk polymers. Performing Monte Carlo (MC)
simulations for a cylindrical body between two parallel plates, we
numerically find the soft elasticity in MC data such that the tensile
stress and strain are consistent with reported experimental
results. Moreover, the elongation is also observed in the results of
MC simulations of a spherical body with free boundaries, and the data
obtained from the MC simulations are also consistent with existing experimental results. 
\end{abstract}

\end{frontmatter}

\section{Introduction}\label{intro}
\begin{figure}[h]
\centering
\includegraphics[width=11.5cm]{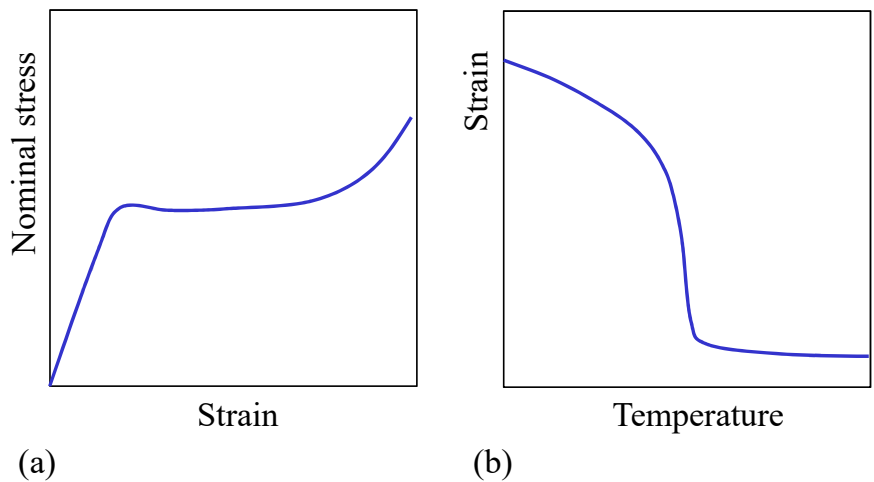}   
\caption{(a) Stress-strain diagram observed in an LCE showing the soft-elasticity characterized by a plateau \cite{Wamer-Terentjev,V-Domenici-2012,Terentjev-JPCM-1999},  and (b) strain $L/L_0$ vs. the temperature $T$ of an LCE film, where $L$ ($L_0$) is the length of the LCE at $T$ (sufficiently high $T$) \cite{V-Domenici-2012,Greve-MacromCP-2001}.
\label{fig-1}
 } 
\end{figure}

The liquid crystal elastomer (LCE), composed of a cross-linked polymer
gel and a liquid crystal (LC), has remarkable properties, such as the
so-called soft elasticity and anisotropic shape transformation (or
elongation)
\cite{Wamer-Terentjev,V-Domenici-2012,Terentjev-JPCM-1999,Greve-MacromCP-2001,K-F-MacMolCP-1998}
(see Figs. \ref{fig-1}(a),(b))). These phenomena are believed to have
intimate connections with the nematic transition of LCs, and this
nematic transition itself is well understood based on the theory of Onsager and/or that of Maier-Saupe \cite{Degenne-Prost-1993,Maier-Saupe-1958}, while the polymer can be described by Flory-Huggins theory \cite{Flory-1959} and the Doi-Edwards' model  \cite{Doi-Edwards-1986}. 
The anisotropy in a $2D$ LCE, which is a membrane, has also been extensively studied, where anisotropic surface constants are assumed in the Hamiltonian for the classical elasticity \cite{Lubensky-PRE-2002,Lubensky-PRE-2003,Xing-Radzihovsky-ANP-2008}.

Soft elasticity is a phenomenon typical of the $3D$ LCE such that the
LCE deforms almost linearly for small stress $\tau$ and considerably deforms for $\tau\!>\!\tau_c$. This large deformation without an increase of $\tau$  creates a plateau in the stress-strain diagram (Fig. \ref{fig-1}(a)).  
This soft elasticity in the $3D$ smectic elastomer has been studied using mean field theory analysis \cite{Stenul-Lubensky-PRL-2005}.

 However, the mechanism for the soft elasticity and elongation is still not fully understood because of the lack of information on the interaction of the LC and the bulk polymer. Although the LC itself and the polymer itself are thoroughly understood as mentioned above, the interplay between them is too complex, and therefore, it remains unclear. 

In this paper, we study the soft elasticity and elongation of an LCE
using  a Finsler geometry (FG) model, and we present the diagrams,
such as the ones in Figs. \ref{fig-1}(a),(b), that are obtained by Monte Carlo (MC) simulations. Although this consistency remains only qualitative, we expect that the FG modeling sheds lights on the unknown mechanism in these anisotropic shape transformations of LCEs. 

The FG model is a coarse-grained model and is defined by extending the Helfrich and Polyakov (HP) model for membranes \cite{HELFRICH-1973,POLYAKOV-NPB1986,Bowick-PREP2001,WIESE-PTCP19-2000,NELSON-SMMS2004,GOMPPER-KROLL-SMMS2004}. This FG model includes a new dynamical variable  $\sigma$. The variable $\sigma$ represents the directional degrees of freedom of LC molecules located at the three-dimensional position ${\bf r}$,  and the non-polar interaction between $\sigma $s is assumed. The polar interaction is also implemented in the model to observe the difference between the polar and non-polar interactions. Using the variables $\sigma$ and  ${\bf r}$, we define 
the Finsler metric,  or in other words, the interaction of this $\sigma$ and the bulk space ${\bf r}$ is directly introduced via the Finsler metric in the Gaussian bond potential $S_1$ \cite{Matsumoto-SKB1975,Bao-Chern-Shen-GTM200,Koibuchi-Sekino-PhysicaA2014}.

The fact that the FG model is an extension of the HP model can be justified as follows: we assume the interaction energy $\lambda S_0$ between $\sigma$s; $S_0$ is  the Lebwohl-Lasher potential \cite{Leb-Lash-PRA1972} (the Heisenberg sigma model energy)  for the case of non-polar (polar) interaction.  For $\lambda\!\to\!0$, the variable $\sigma$ becomes disordered, and consequently, no anisotropy is expected in the system. Therefore, the phase structure of the FG model corresponding to such disordered or isotropic $\sigma$ should be identical to that of the canonical HP model. Indeed, we have verified in the elongation simulation that there is no difference between the stress $\tau$ of the FG model with $\lambda\!=\!0$ and that of the canonical HP model up to a multiplicative constant. Such equivalence was also examined in Ref. \cite{Koibuchi-Sekino-PhysicaA2014}, where the crumpling transition of the $2D$ FG model in the limit of $\lambda\!\to\!0$ is first order as in the HP model  \cite{KD-PRE2002,Kownacki-Mouhanna-2009PRE,Essa-Kow-Mouh-PRE2014}. 

Here, we comment on the difference between the model in this paper and
another variant of the FG model for multicomponent membranes
\cite{Koibuchi-Pol2016}. Several anisotropic shape transformations
(ASTs) are observed in these membranes, and these ASTs are called
domain pattern transitions
\cite{Sarah-PRL2005,Yanagisawa-PRE2010,Lipowsky-SM2009}. For this
special property observed in membranes, the multiplicity of components
is essential as in the low-temperature glasses \cite{GJug-PM2004}. In
the membranes, the origin of ASTs is considered to be the line tension at the domain boundaries \cite{Sarah-PRL2005,Yanagisawa-PRE2010,Lipowsky-SM2009}, and this line tension is understood in the context of FG modeling \cite{Koibuchi-Pol2016}. In this FG model, a new degree of freedom $\sigma$ is also introduced, where $\sigma$ is connected to a scalar function on the surface. This is in sharp contrast to the  model in this paper, where $\sigma$ is a vector. 

The remainder of this paper is organized as follows.  In Subsection
\ref{discrete_model}, the discrete FG model for a $3D$ LCE is
introduced, and the corresponding continuous Gaussian energy and its
discretization technique on the tetrahedrons are described in
Subsection \ref{continuous_model}.  In Section \ref{simulations}, the
partition function on the cylindrical body for calculating the soft
elasticity and that on the spherical body for calculating the
elongation are introduced, and the formula for calculating the tensile
stress is described. Moreover, Section \ref{simulations} also presents Monte Carlo (MC) data for the soft elasticity and the elongation. Finally, in Section \ref{Discussion}, we summarize the results and speculatively comment on future studies along the line of FG modeling.  In Appendix \ref{FG_model}, we briefly introduce the elements of Finsler function and how to obtain the Finsler metric. It is unclear at present whether or not the FG model technique is useful for  MD simulations \cite{Noguchi-JPS2009}.
\section{Model}\label{model}

\subsection{Discrete Model for $3D$ LCE}\label{discrete_model} 
\begin{figure}[h]
\centering
\includegraphics[width=11.5cm]{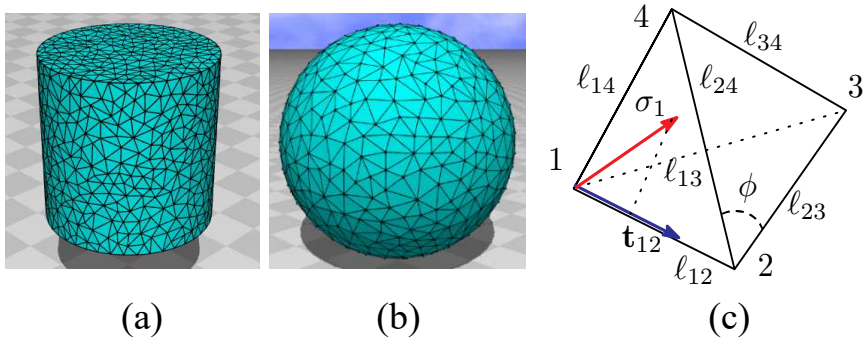}   
\caption{(a) A cylinder of size $(N,N_2)\!=\!(2951,644)$, where the diameter and the height are identical, $N$ is the total number of vertices, and  $N_2$ is the total number of vertices on the side face. (b) A sphere of size  $(N,N_2)\!=\!(2423,1002)$, where  $N$ is the total number of vertices and  $N_2$ is the total number of vertices on the surface.  (c) A tetrahedron on which the discrete Hamiltonian is defined. The variable $\sigma$ is defined at the vertices. 
\label{fig-2}
 } 
\end{figure}

In this subsection, we introduce a discrete FG model for a $3D$ LCE,
which is defined using the technique presented in  Ref. \cite{Koibuchi-Sekino-PhysicaA2014}. 
A cylindrical body and a spherical body in ${\bf R}^3$ are constructed
by the Voronoi tessellation (Figs. \ref{fig-2}(a), (b)) with
tetrahedrons (Fig. \ref{fig-2}(c)), which are composed of vertices, bonds, and triangles. 
A constraint imposed on the lattice construction is that the mean bond length inside the surface is identical to the mean bond length on the surface. 

The discrete Hamiltonian $S({\bf r},\sigma)$ is defined on the tetrahedrons such that
\begin{eqnarray}
\label{Disc-Eneg} 
&& S({\bf r},\sigma)= \lambda S_0(\sigma)+ \gamma S_1({\bf r},\sigma) + \kappa S_2({\bf r}) 
+U_{3D} + U_{2D}, \nonumber \\
&& S_0(\sigma)= \left\{ \begin{array}{@{\,}ll}
                  \sum_{ij}\left( 1-\sigma_i\cdot \sigma_j\right) & \qquad ({\rm polar}) \\
                 \frac{1}{2}\sum_{ij}\left( 1-3(\sigma_i\cdot \sigma_j)^2\right)  & \qquad ({\rm nonpolar}) 
                  \end{array} 
                   \right.,   \nonumber\\
&&S_1=\frac{1}{4\bar N}\sum_{ij}\Gamma_{ij}\ell_{ij}^2, \; \Gamma_{ij}=\sum_{\rm tet}\gamma_{ij}({\rm tet}), \; \ell_{ij}^2=({\bf r}_i-{\bf r}_j)^2,\nonumber\\ 
&&S_2({\bf r})=\sum_{i}\left[1-\cos(\phi_i-\pi/3)\right], \\
&&U_{3D}=\sum_{\rm tet} U_{3D}({\rm tet}),\; 
 U_{3D}({\rm tet})=\left\{ \begin{array}{@{\,}ll}
                 0 &  ({\rm Vol(tet)} >0) \nonumber\\
                 \infty &  ({\rm otherwise}) 
                  \end{array} 
                   \right., \nonumber \\
&&U_{2D}=\sum_{\it \Delta, \Delta^\prime} U_{\it \Delta \Delta^\prime},\; 
 U_{\it \Delta \Delta^\prime}=\left\{ \begin{array}{@{\,}ll}
                 \infty  &  ({\it \Delta, \Delta^\prime} {\rm intersect}) \; \nonumber\\
                 0 &  ({\rm otherwise}) 
                  \end{array} 
                   \right., \nonumber 
\end{eqnarray} 
where ${\bf r}(\in {\bf R}^3)$ is the vertex position and $\sigma_i
(\in S^2\!:\!{\rm unit\; sphere}$) is a variable at vertex $i$. This
$\sigma$ corresponds to the three-dimensional structure of an LC molecule. 
Between the variables $\sigma$, the polar or non-polar interaction is assumed in $S_0$, as mentioned in the Introduction. For the non-polar interaction, the variable $\sigma$ effectively has values on the half sphere ($\sigma_i \in S^2/Z_2$), 
and the direction of $\sigma$ in LC is controlled by the Lebwohl-Lasher potential $\lambda/2\sum_{ij} \left(1\!-\!3(\sigma_i\cdot\sigma_j)^2\right)$ \cite{Leb-Lash-PRA1972}. It is well known that this potential represents the first order transition in LC between the ordered (nematic) and disordered (isotropic) phases. This interaction, defined by Lebwohl and Lasher, is simply called non-polar interaction in the remaining part of this paper. The reason why the polar interaction is assumed in $S_0$  in addition to the non-polar one is to see the difference between the polar and non-polar interactions under the presence of new interaction between $\sigma$ and ${\bf r}$ introduced through the Finsler metric. This interaction of $\sigma$ with ${\bf r}$ plays an essential role for anisotropic shape transformation in our FG model.
Note that the interactions in $S_0$ are not always identical to those assumed for studying the nematic transition, where Landau free energy is always assumed using a mean field for $\sigma$ \cite{Degenne-Prost-1993}.

The $3D$ canonical model is defined without the variable $\sigma$. The Hamiltonian $S$ of the canonical model has almost the same expression of the FG model in Eq. (\ref{Disc-Eneg}) and is given by $S({\bf r})\!=\! \gamma S_1({\bf r}) \!+\! \kappa S_2({\bf r}) \!+\!U_{3D} \!+\! U_{2D}$, where $S_1\!=\!\sum_{ij}\ell_{ij}^2$.

The relation between the discrete expressions of $S_1$ in Eq. (\ref{Disc-Eneg}) and the continuous one is shown in the next subsection. In the discrete Gaussian bond potential $\gamma S_1$ in Eq. (\ref{Disc-Eneg}), the coefficient $\gamma\Gamma_{ij}/4\bar N$ is the effective tension, and 
\begin{eqnarray}
\label{N-bar}
\bar N= \left(1/N_B\right)\sum_{ij} n_{ij}
\end{eqnarray} 
is the mean value of  $n_{ij}$, which is the total number of
tetrahedrons sharing the bond $ij$, and $N_B(=\!\sum_{ij}1)$ is the
total number of bonds. The coefficient $\gamma$ of $S_1$ is always
called surface tension in the case of membranes; however, an LCE is a
$3D$ object, and for this reason, we simply refer to
$\gamma\Gamma_{ij}/4\bar N$ as (microscopic) effective tension. Note that this effective tension has the suffix $ij$, and therefore, it practically plays a role of microscopic string tension of the bond $ij$.  The symbol $\sum_{\rm tet}$ in $\Gamma_{ij}$ denotes the sum over all tetrahedrons sharing bond $ij$. Note that $\bar N$ is a constant that depends on $N$; however, this dependence of $\bar N$ on $N$ disappears at sufficiently large $N$, and therefore, the phase structure of the model is independent of whether $S_1$ is divided by $\bar N$. The reason why the coefficient $1/\bar N$ is included is to remove the multiple contributions of the term $\ell_{ij}^2$ in the sum of $S_1$, where $\ell_{ij}$ is the length of bond $ij$ (see Fig. \ref{fig-2}(c)).  In the $2D$  canonical model for membranes,   $\ell_{ij}^2$ appears only once in the sum of   $S_1\!=\!\sum_{ij}\ell_{ij}^2$, and hence, no extra number is included in $S_1$ \cite{Bowick-PREP2001,GOMPPER-KROLL-SMMS2004}.

The tension $\gamma$ of $S_1$ in Eq. (\ref{Disc-Eneg}) is fixed to $\gamma\!=\!1$. This $S_1$ can also be expressed by the sum over the tetrahedrons $\sum_{\rm tet}$ such that
\begin{eqnarray}
\label{S1-tetra}
S_1({\bf r},\sigma)=\frac{1}{4}\sum_{\rm tet}\left( \gamma_{12}\ell_{12}^2+\gamma_{13}\ell_{13}^2+\gamma_{14}\ell_{14}^2 
+ \gamma_{23}\ell_{23}^2+\gamma_{24}\ell_{24}^2+\gamma_{34}\ell_{34}^2\right), 
\end{eqnarray} 
which is directly obtained from the continuous $S_1$, as we will show in the next subsection. Note that $S_1$ in Eq. (\ref{S1-tetra}) is expressed by the sum of tetrahedrons, while $S_1$ in Eq. (\ref{Disc-Eneg}) is expressed by the sum of bonds, and these are the same and different from each other only in the representation.   The coefficients $\gamma_{ij}(=\!\gamma_{ji})$ in Eq. (\ref{S1-tetra}) are defined by 
\begin{eqnarray}
\label{surface-tension-coeff} 
\gamma_{12}=\frac{v_{12}}{v_{13}v_{14}}+\frac{v_{21}}{v_{23}v_{24}}, \;\gamma_{13}=\frac{v_{13}}{v_{12}v_{14}}+\frac{v_{31}}{v_{32}v_{34}}, \nonumber \\
\gamma_{14}=\frac{v_{14}}{v_{12}v_{13}}+\frac{v_{41}}{v_{43}v_{42}}, \;\gamma_{23}=\frac{v_{23}}{v_{21}v_{24}}+\frac{v_{32}}{v_{31}v_{34}}, \\
\gamma_{24}=\frac{v_{24}}{v_{23}v_{21}}+\frac{v_{42}}{v_{41}v_{43}}, \;\gamma_{34}=\frac{v_{34}}{v_{31}v_{32}}+\frac{v_{43}}{v_{41}v_{42}}. \nonumber
\end{eqnarray} 
These $\{\gamma_{ij}\}$ are a part of $S_1$ and different from  $\gamma(=\!1)$, which is the tension coefficient.   
The variable $v_{ij}$ in Eq. (\ref{surface-tension-coeff}) is the tangential component of $\sigma_i$ along bond $ij$ (Fig. \ref{fig-2}(c))
\begin{eqnarray}
\label{tangential-comp} 
v_{ij}=|{\bf t}_{ij}\cdot \sigma_i|,\quad {\bf t}_{ij}=\vec \ell_{ij}/\ell_{ij}, \quad \vec \ell_{ij}={\bf r}_j-{\bf r}_i,
\end{eqnarray} 
where $v_{ij}\!\not=\!v_{ji}$ in general. Note that 
\begin{eqnarray}
\label{symmetry-gamma} 
\Gamma_{ij}=\Gamma_{ji} 
\end{eqnarray} 
because of the property $\gamma_{ij}\!=\!\gamma_{ji}$.

We comment on the problem that $\gamma_{ij}$ in Eq. (\ref{surface-tension-coeff}) diverges when $v_{ik}$ (or $v_{jk}$) in the denominator becomes $v_{ik}\!\to\!0$ (or $v_{jk}\!\to\!0$), which may happen if $\sigma_i$ (or $\sigma_j$) is vertical to ${\bf t}_{ik}$ (or ${\bf t}_{jk}$).  
To remove this divergence, we introduce a small cut-off $\epsilon\!=\!1\!\times\! 10^{-4}$ or $\epsilon\!=\!1\!\times\! 10^{-5}$ for $v_{ij}$. These values for $\epsilon$ do not make any difference in the results, which are almost identical to those obtained in the limit of $\epsilon\!\to\!0$.  Nevertheless, large values of $\gamma_{ij}$ are actually expected, and this large value of $\gamma_{ij}$ corresponds to a strong repulsive interaction between the tetrahedron edge and $\sigma$. We should emphasize that this strong interaction deforms the shape of tetrahedrons and causes anisotropic shape transformations.

Notably, the interaction between an LC molecule and the bulk polymer is implemented only in the Finsler metric $g_{ab}$ of $S_1$, which will be described in the following subsection. Indeed,  the elements $v_{ij}$ of $g_{ab}$ are defined by both ${\bf r}$ and $\sigma $ as in Eq. (\ref{tangential-comp}). This interaction between ${\bf r}$ and $\sigma $ is complex and cannot simply be expressed as the interaction term in $S_0$ for $\sigma$s; however, this interaction can be understood intuitively as follows. If the direction of $\sigma_i$ changes and its tangential component $v_{ij}$  along bond $ij$ becomes large (small), then the unit of Finsler length ($=\!v_{ij}$) along this bond in the direction from $i$ to $j$ automatically becomes large (small). Consequently, the interaction between vertices $i$ and $j$ described by $S_1$ is effectively influenced by this variation of $v_{ij}$. This influence of $v_{ij}$ on $S_1(\propto \!\sum\gamma_{ij}\ell_{ij}^2)$ is actually understood as follows. From the scale invariance of the partition function $Z$, which will be described in  the next section,  the mean value of $\gamma_{ij}\ell_{ij}^2$ remains constant, while $\gamma_{ij}$ given by Eq. (\ref{surface-tension-coeff}) is locally changeable depending on $v_{ij}$. Hence, the Euclidean bond length $\ell_{ij}$, which is actually connected to the tetrahedron shape, becomes locally changeable depending on $v_{ij}$. 
Thus, the interaction between the LC molecule and the bulk polymer is
coarse grained by this dependence of $S_1$ on $v_{ij}$. This implies that the FG model in this paper is universal in the sense that the interaction is independent of the detailed information on the constituent molecules as in the original HP model for $2D$ membranes.

 The symbol $\phi_i$ in $S_2$ in Eq. (\ref{Disc-Eneg}) is the internal angle of the triangles (see $\phi$ in Fig. \ref{fig-2}(c)), and hence, $\sum_i$ satisfies $\sum_i 1\!=\!3N_T$, where $N_T$ is the total number of triangles. The coefficient $\kappa$ is the rigidity corresponding to the polymer bending stiffness. The potential $U_{3D}$ protects the tetrahedron volume Vol(tet) from being negative.  The self-avoiding potential $U_{2D}$ for the surface is introduced only for the elongation simulations, and it is not used for the soft elasticity simulations.

Here, we comment on the anisotropy expected in the effective tension modulus $\gamma\gamma_{ij}$ included in $\gamma\Gamma_{ij}$ in Eq. (\ref{Disc-Eneg}). The potential force, with respect to the bond potential $\gamma S_1$,  is given by
\begin{eqnarray} 
\label{potential-force}
{\vec f}_i(S_1)=-\gamma \partial S_1/ \partial {\bf r}_i, 
\end{eqnarray} 
which acts on the particle at vertex $i$. Thus, the force along the bond direction $ij$, from  vertex $i$ to  vertex $j$, is given by ${\vec f}_i\cdot {\bf t}_{ij}$, which depends not only on $\ell_{ij}$ but also on $\Gamma_{ij}$, as we observe in the discrete expression of $S_1$ in Eq. (\ref{Disc-Eneg}). For example, the potential force ${\vec f}_1\cdot {\bf t}_{12}$ along the direction of bond $12$ includes the contribution $\gamma\gamma_{12}\ell_{12}$, which comes from the tetrahedron in Fig. \ref{fig-2} (c). Therefore, from the expression $\gamma_{12}$ in Eq. (\ref{surface-tension-coeff}),  we understand that the magnitude of this force $\gamma\gamma_{12}\ell_{12}$ along bond $12$ becomes dependent on $v_{12}$ (and $v_{21}$) and $v_{ij}, (i,j\!\not=\!1,2)$. This implies the possibility that $\gamma\gamma_{12}\ell_{12}$ becomes large (small) compared to $\gamma\gamma_{1j}\ell_{1j}, (j\!\not=\!2)$ if $v_{ij}, (i,j\!\not=\!1,2)$ is  small (large) compared to $v_{12}$ and  $v_{21}$; therefore, it is also possible that the effective tension modulus $\gamma\Gamma_{ij}$  strongly depends on the bond position and the bond direction.

 Note that  a phase transition of $\sigma$ between the ordered and
 disordered phases plays an essential role for the anisotropy in
 $\gamma\Gamma_{ij}$.  The aforementioned dependence of
 $\gamma\Gamma_{ij}$ on the bond position and the bond direction is
 only a local property of $\gamma\Gamma_{ij}$ because the expression
 is given by the local coordinates. Therefore, it is unclear whether
 this property of $\gamma\Gamma_{ij}$ influences the long-distance
 behavior of the model, such as the shape transformation. However, it
 will be clarified that this local property in $\gamma\Gamma_{ij}$
 plays an essential role for the shape transformation. The reason is
 that a global axis by the phase transition of $\sigma$ appears between
 the ordered and disordered phases.  Along this axis that appeared spontaneously,  the variable $\sigma$ aligns, and therefore, the value of $\gamma\Gamma_{ij}$ along this axis becomes different from those for other directions, and for this reason, the anisotropic shape transformation emerges.
 
\subsection{Continuous Gaussian energy and the discretization}\label{continuous_model}
The continuous Gaussian bond potential $S_1$ is a $3D$ extension of
the Hamiltonian of the $2D$ model for membranes
\cite{WHEATER-JP1994}. The $2D$ model for membranes  is considered to be a natural extension of Doi-Edwards' model for linear polymers \cite{Doi-Edwards-1986}, as mentioned in \cite{Bowick-PREP2001}. We note that the $3D$ extension of $S_1$ from the $2D$ model is straightforward because  the expression of $S_1$ for the $3D$  model is exactly identical to  the expression of $S_1$ for the $2D$  model in Ref.\cite{Koibuchi-Pol2016}, except for the parameters $x_a, (a\!=\!1,\cdots,D)$, where $D\!=\!2$ or $D\!=\!3$. The expression  of $S_1$ is given by
\begin{eqnarray} 
\label{cont_S}
S_1=\int \sqrt{g}d^3x g^{ab} \frac{\partial {\bf r}}{\partial x_a}\cdot \frac{\partial {\bf r}}{\partial x_b}, 
\end{eqnarray} 
where the LCE position ${\bf r}$ is considered to be a mapping from a three-dimensional parameter space $M$ to ${\bf R}^3$ in the HP prescription, $g^{ab}$ is the inverse of the Finsler metric $g_{ab}$, and $g$ is its determinant. 
The discrete version of this $g_{ab}$ is given by 
\begin{equation} 
\label{Finsler_metric}
g_{ab}=\left(  
       \begin{array}{@{\,}ccc}
         1/v_{12}^2  & \;  0             & \; 0 \\
               0     & \; 1/v_{13}^2     & \; 0 \\
               0     & \; 0              & \; 1/v_{14}^2  \\
       \end{array} 
       \\ 
 \right),
\end{equation}
where $\{v_{ij}\}$ are the same as those used in
Eq. (\ref{surface-tension-coeff}). The reason why we call this metric "discrete" is because $v_{ij}$ is defined on the tetrahedrons as shown in Eq. (\ref{tangential-comp}).  
This $g_{ab}$ in Eq. (\ref{Finsler_metric}) is obtained from the Euclidean metric $\delta_{ab}$ by replacing the diagonal elements with $1/v_{ij}^2$, which reflects an asymmetry in the direction of $\sigma$ \cite{Koibuchi-Sekino-PhysicaA2014}.  Thus, the asymmetry is introduced in the model via the asymmetrical quantity $v_{ij}$. The phrase "$v_{ij}$ is asymmetric" means that "$v_{ij}\!\not=\!v_{ji}$ in general". 

The  integration and the partial derivatives of ${\bf r}$ in $S_1$ of Eq. (\ref{cont_S}) are replaced by
\begin{eqnarray} 
\label{discretization}
&&\int \sqrt{g}d^2x \to \sum_{\it \Delta} v_{12}^{-1}v_{13}^{-1}v_{14}^{-1}, \nonumber \\
&&\partial_1  {\bf r} \to {\bf r}_2- {\bf r}_1, \;
\partial_2  {\bf r} \to  {\bf r}_3- {\bf r}_1, \;
\partial_3  {\bf r} \to  {\bf r}_4- {\bf r}_1
\end{eqnarray} 
 on the tetrahedron in Fig. \ref{fig-2}(c), where the local coordinate origin is assumed at vertex $1$. Because we have four possible coordinate origins in a tetrahedron, summing over all possible forms of $\partial_a{\bf r}$, we obtain $S_1$ in Eq. (\ref{S1-tetra}).
This summation of the possible coordinate origins for $S_1$ makes
$\gamma_{ij}$ and hence $\Gamma_{ij}$ symmetric under the exchange of
$ij$, as shown in Eq. (\ref{symmetry-gamma}). This symmetry implies
that the effective tension modulus $\gamma\gamma_{ij}$ is also
symmetric. We must note that $\gamma_{ij}$ depends on the direction
($\Leftrightarrow \gamma_{ij}\not=\!\gamma_{ik}$ if $j\!\not=\!k$),
although it is symmetric ($\gamma_{ij}\!=\!\gamma_{ji}$); symmetric and isotropic are not always the
same. This symmetry in $\gamma_{ij}$ and $\Gamma_{ij}$ arises from the
fact that the variable $\sigma$  is defined on the vertices.  This is
in sharp contrast to the case in which the elements of $g_{ab}$ are
defined on the triangles, where $\gamma_{ij}\!\not=\!\gamma_{ji}$ in
general \cite{Koibuchi-Pol2016}. Therefore, such a Finsler geometry
model, in which $g_{ab}$ is defined on the triangles (or tetrahedrons), forms another interesting class of models, as mentioned in the Introduction. 

\section{Monte Carlo Simulations}\label{simulations}
The Metropolis MC technique is used for the update of ${\bf r}$ to ${\bf r}^\prime\!=\!{\bf r}\!+\!\delta{\bf r}$, where $\delta{\bf r}$ is a random vector inside a small sphere \cite{Mepropolis-JCP-1953,Landau-PRB1976}.  The radius of this sphere is fixed such that the acceptance rate for ${\bf r}$ approximately equals $50\%$.  The variable $\sigma$ is updated to $\sigma^\prime (\in S^2)$, which is randomly defined with three different random numbers, and therefore, it becomes independent of the current $\sigma$. In this update, the rate of acceptance is very low, at least for large $\lambda$; however, we have no problem on the convergence because the convergence rate of $\sigma$ is supposed to be very rapid compared to that of ${\bf r}$. The rapid convergence of $\sigma$ arises from the fact that the phase space volume of $\sigma$ is finite ($S^2$ unit sphere) while that of ${\bf r}$ is infinite (${\bf R}^3$). $N$ updates of  ${\bf r}$ and $N$ updates of $\sigma$ are called one Monte Carlo sweep (MCS). 
For the soft elasticity simulations, the data measurements are
executed at every $1000$ MCS during $5\!\times\! 10^7$ to $2\!\times\!
10^8$ MCS after the $5\times 10^6$ thermalization MCS.  Additionally,
for the elongation simulations, a relatively small number of MCS is
sufficient, although the spherical body is not supported by any
boundaries. The thermalization MCS is $2\!\times\! 10^6$, and the MCS
for measurements is  $4\!\times\! 10^7$ to $1\!\times\! 10^8$ after
the thermalization MCS. The reason for such a small number of MCS is
because the deformation of a $3D$ body is very small compared to the case of $2D$ membranes, where the small deformation means that the practical phase space volume for ${\bf r}$ in ${\bf R}^3$ becomes very small compared to the one for ${\bf r}$ of $2D$ membranes.

\subsection{Soft elasticity}\label{soft-simu}
Soft elasticity is experimentally observed for LCE films under small external forces \cite{Wamer-Terentjev,V-Domenici-2012,Terentjev-JPCM-1999} (see Fig \ref{fig-1}(a)). To observe the soft elasticity in our model, we use a cylinder of size $(N,N_B,N_T,N_{\rm tet})\!=\!(4255,29303,48860,23811)$, where $N_B,N_T,N_{\rm tet}$ denote the total number of bonds, triangles, and tetrahedrons, respectively. It is easy to verify that $N\!-\!N_B\!+\!N_T\!-\!N_{\rm tet}\!=\!1$ for any simply connected volume discretized by tetrahedrons. 

On the upper and lower faces, which are round disks in the initial configuration (Fig. \ref{fig-2}(a)), the vertices are allowed to move on the faces  except a vertex of each boundary surface. The reason why these two vertices (one on the upper surface and the other on the lower surface) are fixed is to protect the vertices of the boundary surfaces from moving freely in the horizontal direction. This is only for the FG models, and no constraint is imposed on the canonical model (which is simulated and only snapshots are shown). In the FG models, we choose these two vertices of which the distance from the $z$-axis  at  the cener of the initial boundary disks is the minimum. We should note that the vertices including these two ones on the boundary surfaces are allowed to move into the $z$-direction for small distance. This will be described below in more detail. 

The self-avoidance for the tetrahedrons, implemented by $U_{3D}$ in Eq. (\ref{Disc-Eneg}), also prohibits the triangles from folding on these upper and lower faces. However, the surface self-avoiding interaction is neglected on the side face of the cylinder, as mentioned in Section \ref{model}; consequently, the triangles can intersect on the side face. However, this self-intersection is expected to be negligible because the height of the cylinder is fixed during the simulations and the fluctuation of the side face is considerably suppressed.

Let $N_{2d}$ be the total number of vertices on the upper and lower faces, and let $N_{3d}$ be defined by
\begin{eqnarray} 
\label{no-of-vertioces}
N_{3d}=N-N_{2d}.
\end{eqnarray} 
We have  $N_{2d}\!=\!502$ and  $N_{3d}\!=\!3723$  for the cylinder of size $N\!=\!4225$ used for the soft elasticity simulations in this paper. 
Therefore, the partition function of the model on the cylinder is given by 
\begin{eqnarray}
\label{part-func-cyl} 
&&Z_{\rm cyl}=\sum_\sigma \int \prod_{i=1}^{N_{2d}}d{\bf r}_i \prod_{i=1}^{N_{3d}}d{\bf r}_i \exp[-S({\bf r},\sigma;L)],  \\ 
&& S({\bf r},\sigma; L)= \lambda S_0(\sigma)+ \gamma S_1({\bf r},\sigma) + \kappa S_2({\bf r})+U_{3D}  + U_{B}, \nonumber \\
&&
U_{B}= \sum_{i\in {\rm boundary}} U_{B}({\bf r}_i),\;  U_{B}({\bf r}_i)=\left\{ \begin{array}{@{\,}ll}
                 \infty &  (|z_i-L| >\delta_B\; {\rm or}\; |z_i| >\delta_B) \nonumber\\
                  0 &  ({\rm otherwise}) 
                  \end{array} 
                   \right., \nonumber             
\end{eqnarray}
where  $S({\bf r},\sigma;L)$ denotes that the height of the cylinder is fixed to $L$ (which should not be confused with the Finsler function in Appendix \ref{FG_model}). The symbols $\int\prod_{i=1}^{N_{2d}}d{\bf r}_i$ and $\int \prod_{i=1}^{N_{3d}}d{\bf r}_i$ in $Z_{\rm cyl}$ denote the $2N_{2d}$ and $3N_{3d}$-dimensional multiple integrations. Note that $N_{2d}$ in $\int\prod_{i=1}^{N_{2d}}d{\bf r}_i$ should be replaced by $N_{2d}\!-\!2$ because of the fixed points mentioned above. However, we neglect the number $-2$ in this expression.   

 As mentioned in Section \ref{discrete_model},   the variables $\sigma_i$ on the upper and lower boundaries are strongly influenced by the boundary bonds if the surfaces are flat, because $v_{ij}\!=\!|{\bf t}_{ij}\cdot \sigma_i|$ becomes zero for  $\sigma_i$ that is vertical to the flat boundary surfaces. Therefore the variables $\sigma_i$ can not be vertical to the flat boundary surfaces.  To remove such unnatural effect, we introduce the potential $U_B$ assuming  that the vertices on the boundary surfaces can move into the height direction within $\delta_B$ from the boundary surfaces at the height positions $z\!=\!0$ (lower) and $z\!=\!L$ (upper). The small height $\delta_B$ is fixed to the half of the mean bond length of the initial cylinder configuration, where the height and diameter of the cylinder are given by $L$ and $D_0$, respectively. From this definition, we have
\begin{eqnarray} 
\label{small-height}
\frac{\delta_B }{L}\left(=\frac{\rm mean\; bond\; length}{\rm height\;of\;cylinder}\right) \to 0 \quad (N\to\infty), 
\end{eqnarray} 
because the size of tetrahedron is independent of $N$ while the height $L$ increases with increasing $N$. This implies that the influence of $U_B$ is negligible for sufficiently large $N$.

From the scale invariance of $Z_{\rm cyl}$ under the transformation ${\bf r} \!\to\!\alpha {\bf r}$, where $\alpha$ is the scale parameter, we have  \cite{WHEATER-JP1994}
\begin{eqnarray}
\label{scale-inv-of-Z} 
\partial \log Z_{\rm cyl}(\alpha)/\partial \alpha |_{\alpha=1}=0.
\end{eqnarray}

Since $L$ remains unchanged under the scale transformation ${\bf r} \!\to\!\alpha {\bf r}$, we have the expression 
\begin{eqnarray}
\label{scaled-Z} 
Z_{\rm cyl}(\alpha)=\alpha^{2N_{2d}+3N_{3d}} \sum_\sigma \int \prod_{i=1}^{N_{2d}}d{\bf r}_i \prod_{i=1}^{N_{3d}}d{\bf r}_i \exp[-S(\alpha{\bf r},\sigma;\alpha^{-1}L)], 
\end{eqnarray}
where $\alpha^{-1}L$ (rather than $L$) should be remarked. 
Because of this $L$ dependence, the left-hand side  $\partial \log
Z_{\rm cyl}(\alpha)/\partial\alpha$ of Eq. (\ref{scale-inv-of-Z})
should include the term of the partial derivative with respect to
$\alpha^{-1}L$ in $\exp[-S(\alpha{\bf r},\sigma;\alpha^{-1}L)]$ of
Eq. (\ref{scaled-Z}). To clarify this point, we temporarily write $Z_{\rm cyl}(\alpha)$ as $Z_{\rm cyl}(\alpha; \alpha^{-1}L)$. Thus, we have  
\begin{eqnarray}
\partial Z_{\rm cyl}(\alpha; \alpha^{-1}L)/\partial\alpha =\partial Z_{\rm cyl}(\alpha; *)/\partial\alpha+\partial Z_{\rm cyl}(*; \alpha^{-1}L)/\partial\alpha, 
\end{eqnarray}
where $\partial Z_{\rm cyl}(\alpha; *)/\partial\alpha$ and $\partial Z_{\rm cyl}(*; \alpha^{-1}L)/\partial\alpha$ denote the partial derivatives with respect to $\alpha$ except $\alpha$ in  $*$. This is only the Leibniz rule. Let us write $Z_{\rm cyl}(*; \alpha^{-1}L)$ simply as $Z_{\rm cyl}(\alpha^{-1}L)$; then, using the relation  
\begin{eqnarray}
\label{differential-rule} 
&&\partial Z_{\rm cyl}(\alpha^{-1}L)/\partial\alpha\nonumber \\
=&&[\partial Z_{\rm cyl}(\alpha^{-1}L)/\partial(\alpha^{-1}L)][\partial(\alpha^{-1}L)/\partial\alpha]\nonumber \\
=&&-L\alpha^{-2}\partial Z_{\rm cyl}(L)/\partial L,
\end{eqnarray}
we have 
\begin{eqnarray}
\label{scale-inv} 
2\gamma\langle S_1\rangle-2{N_{2d}}-3N_{3d}=-(L/Z_{\rm cyl})\partial Z_{\rm cyl}(L)/\partial L.
\end{eqnarray}
On the left-hand side of Eq. (\ref{scale-inv}), the terms
$2\gamma\langle S_1\rangle$ and $2N_{2d}+3N_{3d}$ come
from the partial derivatives of $S_1(\alpha{\bf r})\!=\!\alpha^2
S_1({\bf r})$ and $\alpha^{2N_{2d}+3N_{3d}}$, respectively, in $Z_{\rm cyl}(\alpha)$ of Eq. (\ref{scaled-Z}). 
To calculate the right-hand side of Eq. (\ref{scale-inv}), we assume
that the cylinder is a continuous elastic object. Therefore, the free energy $F(L)$ of this elastic cylinder is given by
\begin{eqnarray}
\label{free-energy} 
F(L)=\int_{L_0}^L \frac{f^2}{EA} dz,
\end{eqnarray}
where $f$ is the external tensile force applied to the cylinder along
the height (or $z$) direction and $E$ and $A$ are the Young's
modulus and the sectional area perpendicular to the $z$ direction, respectively. Thus, using the relation 
\begin{eqnarray}
\label{free-energy-def} 
F(L)=-\log Z_{\rm cyl} \quad \left[\Leftrightarrow  Z_{\rm cyl}=\exp -F(L)\right],
\end{eqnarray}
we have from the right hand side of Eq. (\ref{scale-inv}) that
\begin{eqnarray}
&&-(L/Z_{\rm cyl})\partial Z_{\rm cyl}(L)/\partial L=L\partial F(L)/\partial L\nonumber \\
&&=(f^2/EA)L=(f/A)^2 AL/E.
\end{eqnarray}
In the final term, $f/A$ is understood to be the true stress. Therefore, by replacing $A$ with $A_0$, we obtain the nominal stress $\tau(=\!f/A_0)$ such that
\begin{eqnarray}
\label{stress} 
\tau(L)=\sqrt{\left(2\gamma\langle S_1\rangle-2N_{2d}-3N_{3d}\right)E\,/\,{LA_0}},
\end{eqnarray}
where $\gamma\!=\!1$, $E\!=\!1$  and $A_0$ defined by
\begin{eqnarray}
\label{def-of-area} 
A_0=\langle V_0\rangle/L_0
\end{eqnarray}
is the sectional area of the cylinder of height $L_0$ and volume $\langle V_0\rangle$. This cylinder causes $\tau\to 0$ because the height $L_0$ satisfies $2\gamma\langle S_1\rangle\!-\!2N_{2d}\!-\!3N_{3d}\!=\!0$. 
Note that the true stress is obtained by replacing $LA_0$ with $V(=\langle V\rangle)$ in Eq. (\ref{stress}).

 We must emphasize that the formula for $\tau(L)$ in Eq. (\ref{stress}) is obtained from Eq. (\ref{scale-inv}) under the assumption that $Z_{\rm cyl}(L)$ in the right hand side is given by the macroscopic free energy $F(L)$ in Eq. (\ref{free-energy}). In contrast, all quantities in the left hand side of Eq. (\ref{scale-inv}) are given by the microscopic mechanics, which is defined by  the discrete Hamiltonians and the partition function. All microscopic data such as $\lambda$, $\gamma$ and $\kappa$,  assumed for the calculation of the left hand side including $\langle S_1\rangle$, are connected to the macroscopic quantity by this formula in Eq. (\ref{scale-inv}). Note that the macroscopic shear modulus is not included in $F(L)$, because the present calculation is limited only for tensile deformation of LCE. In fact, to obtain the macroscopic shear stress of LCE in our formulation, we have to define the free energy corresponding to shear deformations, and this is out of the scope of this paper.

The initial configuration of $\sigma$ for the simulations of the non-polar model is assumed to be radially symmetric \cite{V-Domenici-2012,Terentjev-JPCM-1999} such that 
\begin{eqnarray}
\label{initial-cond-nonpolar} 
(\sigma_x,\sigma_y,\sigma_z)=(\cos \theta, \sin\theta,0), \quad ({\rm nonpolar}),
\end{eqnarray}
where $\theta$ is the polar angle of the vertex position ${\bf r}$ on
the plain perpendicular to the $z$ axis. A random start is also performed for the non-polar model to see the dependence on the initial configurations. 
In contrast, only the random configuration of $\sigma$ is assumed for the polar model because the radially symmetric configuration is unstable at the center $x\!=\!y\!=\!0$ of the cylinder in this case. 

As mentioned above, the  initial height $L_0$ of the cylinder is fixed such that we have 
\begin{eqnarray}
\label{initial-cond} 
\gamma\langle S_1\rangle=N_{2d}+(3/2)N_{3d}, 
\end{eqnarray}
which corresponds to the case of $\tau\!=\!0$ from
Eq. (\ref{stress}). This initial cylinder height $L_0$ for
$\tau\!=\!0$ depends on the parameters $\lambda$ and $\kappa$, as well
as on whether the model is polar or non-polar. In the simulations for
$\tau\!=\!0$, the mean value of the diameter becomes different from  $L_0$ in general even though the cylindrical lattice is constructed such that the height equals the diameter.

\begin{figure}[h]
\centering
\includegraphics[width=9.5cm]{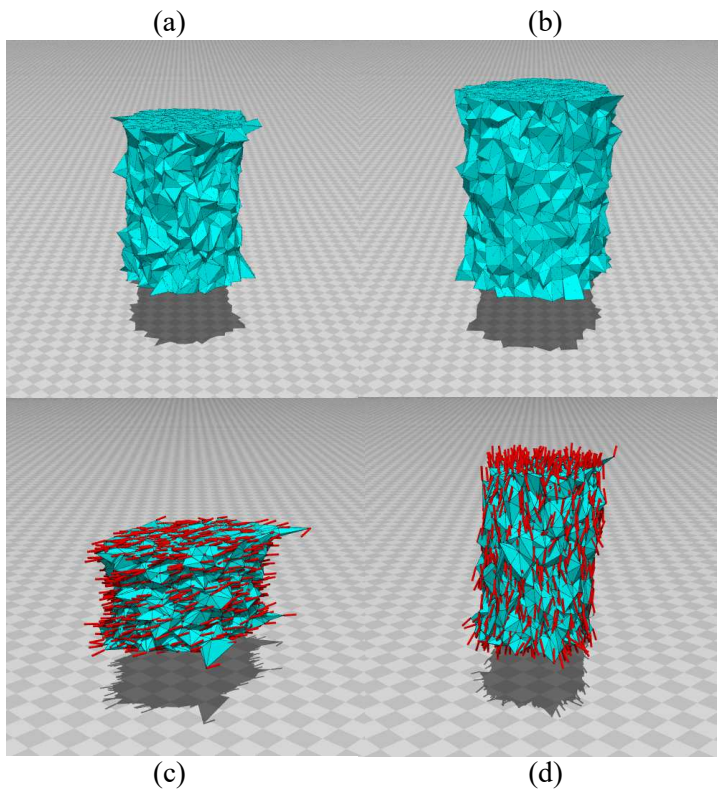}  
\caption{Snapshots of the soft elasticity simulations of the canonical model for (a) $L/L_0\!=\!1.49$, $\kappa\!=\!0$ and (b) $L/L_0\!=\!1.61$, $\kappa\!=\!0.4$, and the non-polar FG model for (c) $L/L_0\!=\!1.68$ and (d) $L/L_0\!=\!1$ with $\lambda\!=\!1$, $\kappa\!=\!0.3$. $N\!=\!4255$. }
\label{fig-3} 
\end{figure}
First, we show the snapshots of the canonical model in Figs.\ref{fig-3} (a), (b). The canonical model corresponds to the FG model with $\lambda\!=\!0$. 
In the canonical model simulations, the upper/lower boundary vertices are allowed to move freely in the horizontal plain, and therefore, if the lattice construction is not uniform, it is suspected that the lattice unexpectedly deforms. These snapshots indicate that the height direction of the cylinder does not deviate from the $z$-direction. This implies that the lattice is suitably constructed for our purpose.

The snapshots in Figs. \ref{fig-3} (c), (d) are those of the non-polar model for  $\lambda\!=\!1$, $\kappa\!=\!0.3$ with the strain (c) $L/L_0\!=\!1$ and (d) $L/L_0\!=\!1.68$, where the ordered initial configuration is assumed. The variable $\sigma$ defined at the vertices is represented by the small cylinders (or rods). We find that $\sigma$ almost completely aligns along the horizontal and height directions in (c) and (d), respectively. This implies that $\sigma$ changes its direction according to the applied external tensile force. 
 The surface is relatively rough, however, we should emphasize that the configurations are  connected with not only macroscopic mechanics but also microscopic mechanics. Indeed, we assume two different partition functions in Eqs. (\ref{part-func-cyl}) and (\ref{free-energy-def}).

\begin{figure}[h]
\centering
\includegraphics[width=11.5cm]{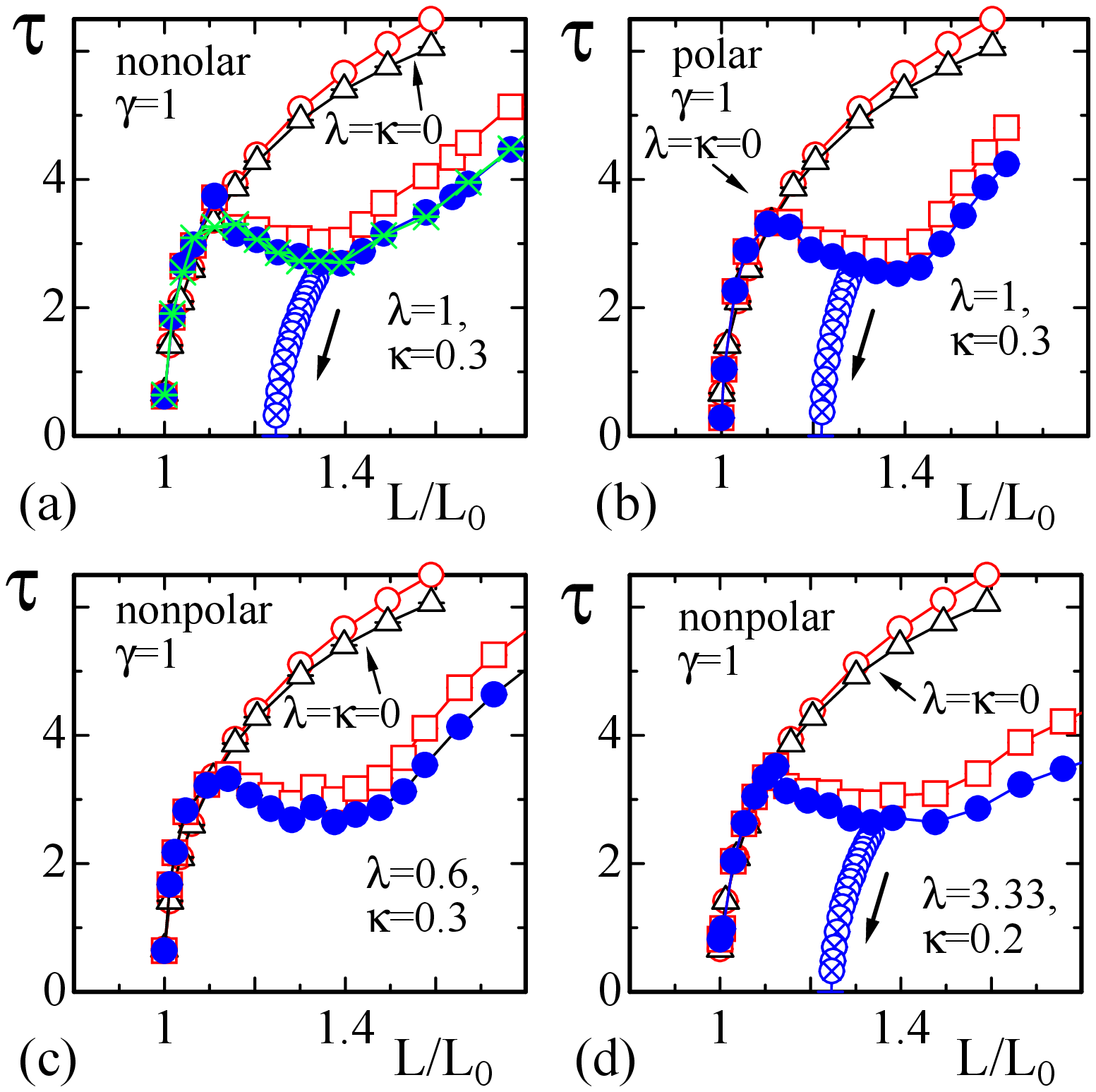}  
\caption{The nominal stress $\tau$ (${\color{blue} \bullet}, \triangle$) vs. strain $L/L_0$ (a) non-polar, (b) polar and (c),(d) non-polar cases. The corresponding true stresses (${\color{red} \bigcirc}, {\color{red} \square}$) are also shown. Every figure includes the symbols (${\color{red} \bigcirc}, \triangle$), which are the data for $\lambda\!=\!0$ and $\kappa\!=\!0$, corresponding to the model without the variable $\sigma$. The symbols (${\color{blue} \otimes}$), guided by an arrow, are the results of hysteresis simulations, in which the variable $\sigma$ is kept frozen to the configuration of the expanded cylinder. In (a), the symbols (${\color{blue} \bullet}, {\color{red} \square}$) denote the ordered start configuration for the variable  $\sigma$, while the symbol (${\color{green} \times}$), the nominal stress, denotes the random start configuration.  The curve $\tau$ vs. $L/L_0$ has plateaus like in the experimental data reported in \cite{Wamer-Terentjev,V-Domenici-2012,Terentjev-JPCM-1999}. 
\label{fig-4}
} 
\end{figure}
The symbols ($\triangle$) in Fig. \ref{fig-4} are obtained under
$\lambda\!=\!\kappa\!=\!0$. For $\lambda\!=\!0$, the variable $\sigma$
and hence $\gamma_{ij}$ in Eq. (\ref{surface-tension-coeff}) becomes
random. In this case, we have no difference (up to a multiplicative
factor) between  $\tau$ ($\triangle$) vs. $L/L_0$ for $\lambda\!=\!0$
and  that of the canonical model without the variable $\sigma$, where $S_1$ is
the ordinary one $S_1\!=\!\sum_{ij}\ell_{ij}^2$. This smooth behavior
without the cusp in the curve of $\tau$ vs. $L/L_0$ remains almost
unchanged for non-zero $\kappa$ if $\lambda\!=\!0$. In contrast, the results for $\lambda\!=\!\kappa\!=\!0$ are quite different  from those (${\color{blue} \bullet}$)  for large $\lambda$, such as $\lambda\!=\!0.6$, $\lambda\!=\!1$ and  $\lambda\!=\!3.33$ with non-zero $\kappa$, where $\tau$ has a plateau with a cusp. 

We find that there is no dependence of the results on the initial configurations for the variable $\sigma$ in the non-polar model. Indeed, the nominal stresses (${\color{green} \times}$) in Figs. \ref{fig-4}(a) obtained by random start MCs are almost identical to those (${\color{blue} \bullet}$) obtained by the ordered start MCs. The linear behavior of $\tau$ can be seen for small $\tau$ (at least in the region $\tau<2$) in both polar and non-polar models.  

Young modulus $E$ can be evaluated by Eq.(\ref{stress}). Recalling that $\tau(L;E\!=\!1)$ corresponds to the numerical data $\tau$, we can evaluate $E$ by $E\!=\!\left(\tau_{\rm exp}/\tau\right)^2$ (see Appendix \ref{formula-for-tau}). For $\tau_{\rm exp}$ in this expression $E\!=\!\left(\tau_{\rm exp}/\tau\right)^2$, we use the experimental data reported in Refs.\cite{Wamer-Terentjev,V-Domenici-2012,Terentjev-JPCM-1999}, which approximately range from $\tau_{\rm exp}\! =\! 10^3 [{\rm Pa}]$ to  $\tau_{\rm exp}\!=\!10^4 [{\rm Pa}]$.  Thus, we have approximately $E\!=\! 0.001[{\rm GPa}] \sim E\!=\!0.1 [{\rm GPa}]$, which are comparable to experimental data of elastomers \cite{Herrmann-PolTes2012}.

Note that the length of the plateau depends on $\kappa$ (Fig. \ref{fig-4}(c)). When both $\lambda$ and
$\kappa$ are large, such that $\lambda\!=\!3.33$ and $\kappa\!=\!1$  for
example, $\tau$ discontinuously changes (this is not plotted). More
precisely, $\tau$ is discontinuously reduced if $L/L_0$ increases, and
subsequently, $\tau$ begins to increase with increasing $L/L_0$. At
this transition point, a discontinuous change can also be observed in
the volume. This result implies that the transition changes from the second to the first order if $\kappa$ ($\lambda$) is increased for sufficiently large $\lambda$ ($\kappa$). We also note that the shape of $\tau$ vs. $L/L_0$ with the plateau is close to the experimental data of the stress-strain curve in \cite{V-Domenici-2012,Terentjev-JPCM-1999} (see Fig. \ref{fig-1}(a)).  It is also confirmed that the plateau shape is almost independent of whether the interaction is polar or non-polar. The true stress (${\color{red} \square}$) is almost the same as the nominal stress (${\color{blue} \bullet}$) for all combinations of $\lambda$ and $\kappa$. 

Now, we comment on the boundary condition for the upper/lower surface of the FG models. As we have seen in the snapshot of  Fig.\ref{fig-3}(d), the axis of the cylindrical body slightly deviates from the vertical direction. The deviation of the axis from the vertical direction is relatively small in the configurations for $\kappa\!=\!0.3$, however, slightly larger deviation  can be seen in some of the snapshots for small $\kappa$ such as $\kappa\!=\!0.1$ and even for $\kappa\!=\!0.2$, although a vertex on the upper/lower surface is fixed  to prevent the surface vertices from moving into the horizontal direction. For this reason,  we examine another condition for the upper/lower surface such that the center of mass of the vertices is fixed (inside a small circle of which the radius is given by the mean bond length) on the horizontal plain to keep the axis vertical to the plain. As a consequence, the axis always becomes vertical to the horizontal plain independent of the value of $\kappa$. However, the obtained results are different from those  plotted in Fig. \ref{fig-4}. Indeed, $\tau$ has a discontinuity just like the one observed for large $\lambda$ and $\kappa$ under the first boundary condition as mentioned above. This implies that the latter boundary condition is more effective than the former one to prevent the surface vertices from moving horizontally. Moreover, this indicates that the shear stress should also be considered in the latter boundary condition for relatively small region of $\kappa$, however, we do not go into detail of this problem in this paper.

 The strong influence of $\sigma$ on the edge direction of tetrahedron is not always limited to the boundary surfaces, it is also expected everywhere inside the body. Therefore, if the variables $\sigma_i$ align along a direction, which is determined spontaneously or by some external forces,  the tetrahedrons deform to protect $\sigma_i$ being vertical to the bond $ij$. This is an explanation of the mechanism of shape deformation in LCE in the context of FG modeling.

Here, we note that our result indicates the possibility that the
plastic deformation of metallic materials is partly understood along
the context of Finsler geometry modeling. The stress-strain curve of
these metals  has a plateau, which represents the plastic deformation,
just like the one in $\tau$ vs. $L/L_0$ in Fig. \ref{fig-4}. This
implies that the plastic deformation shares a common origin with the
soft elasticity. Indeed, the work-hardening phenomenon can be observed
in our model. The stress $\tau$ decreases as $L/L_0$ decreases
(Fig. \ref{fig-4} (${\color{blue} \otimes}$)),  which are obtained by
the hysteresis simulations (but no hysteresis because of the rapid
convergence). In these simulations, the variable $\sigma$ is kept
frozen to the configuration of the expanded cylinder, just like a
glass \cite{GJug-2015rview}.  The hysteresis simulations consist of several consecutive simulations in which the final configuration of ${\bf r}$ is used as the initial one of the next simulation, where  $L$ is decreased step by step in this case. If $\sigma$ is not kept frozen and treated as a dynamical variable in the hysteresis simulations with decreasing $L$, the stress $\tau$ returns along the same position of the original $\tau$.

Additionally, note that a plateau can also be observed in the stress-strain curves of porous or cellular materials. In cellular materials, the structural change of the cellular sections is reflected in the plateau of the stress-strain curve. This structural change in the cellular sections is connected with a rotational symmetry, and therefore, the plateau can also reflect a transition from rotationally symmetric to non-symmetric states. Note that the reverse transition is not always observed in cellular solids \cite{Gibson-CMMM2010}.

\begin{figure}[h]
\centering
\includegraphics[width=11.5cm]{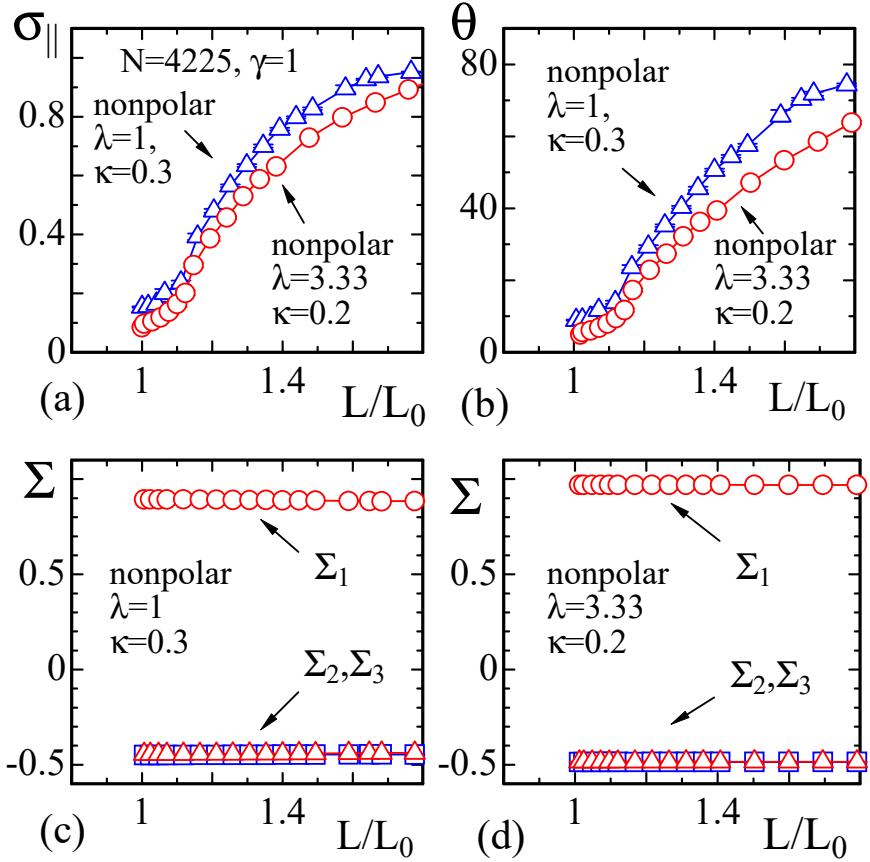}  
\caption{(a) $\sigma_{||}=\sum_i|\sigma_{iz}|/N$ vs. $L/L_0$, (b) the angle $\theta$ vs. $L/L_0$, where $\theta\!=\!90\!-\!(180/\pi) \cos^{-1}\sigma_{||}$, and the eigenvalues $\Sigma_{1,2,3}$ vs. $L/L_0$ for (c) $\lambda\!=\!1$,  $\kappa\!=\!0.3$  and (d) $\lambda\!=\!3.33$,  $\kappa\!=\!0.2$. 
\label{fig-5}
 } 
\end{figure}
The orientation of $\sigma$ along the height direction can be reflected in 
\begin{eqnarray}
\label{sigma-parallel-def} 
\sigma_{||}=(1/N)\sum_i |(\sigma_i)_z| 
\end{eqnarray}
vs. $L/L_0$ in Fig.\ref{fig-5}(a). We observe that 
\begin{eqnarray}
\label{sigma-parallel} 
 \sigma_{||}\to 1 \;\; (\sigma_{||}\to0)\quad  L/L_0 >> 1 \;\;(L/L_0\to1).
\end{eqnarray}

 Therefore, this directional change
 in $\sigma$ is understood to be a structural change, which undergoes
 a phase transition. Note that this change in $\sigma$ is reflected in
 the external mechanical properties, such as $\tau$. Indeed, we
 observe that  $\sigma_{||}$ discontinuously changes as a function of
 $L/L_0$ for sufficiently large $\lambda$ and $\kappa$   (this is not
 plotted). Thus, the discontinuous change of $\sigma$ can be reflected
 in $\tau$ as its discontinuous change. At this discontinuous
 transition point, the volume and the sectional area also
 discontinuously change as mentioned above, although these
 discontinuities are relatively small because the cylindrical shape
 remains identical.  This structural change is considered to be very
 similar to the liquid-solid (or liquid-vapor) phase transition of
 materials if the curve of $\tau$ vs. $L/L_0$ in Fig. \ref{fig-4} is
 identified with the pressure vs. density curve of materials.  Indeed,
 the pressure increases with increasing density in the pressure
 vs. density curve of the liquid-vapor transition. In the small  density region, where the material is in the vapor phase,  the pressure is expected to increase almost linearly with the density. As the density increases further, the pressure stops increasing and has a plateau, where the material turns to be in the two-phase coexistence state. If the density is further increased, the plateau of the pressure terminates, and the pressure rapidly increases. In this case, a real structural change occurs in the material, whereas in the LCE model, the terminology "structural change" only refers to the change in $\sigma$.

In Fig. \ref{fig-5}(b), we plot the mean value of the angle $\theta$ defined by 
\begin{eqnarray}
\label{angle-theta} 
\theta=90-(180/\pi) \cos^{-1}\sigma_{||}.
\end{eqnarray}
 The plotted results show that the angle increases from $\theta\!\simeq \!0$ to  $90$ degree with increasing $L/L_0$. However, $\theta$ becomes not exactly zero ($90$)  degree even for small (large) strain $L/L_0$. This is not always consistent to the actual experimental data reported in  \cite{V-Domenici-2012,K-F-MacMolCP-1998}, where the scaled strain is used instead of $L/L_0$. The reason for these deviations in the data is that the tetrahedron hardly deforms for relatively large $\kappa$, and this protects $\sigma$ from being parallel to the $z$-axis. 
The reason why  $\theta$ for $\kappa\!=\!0.2$ is slightly smaller than $\theta$ for $\kappa\!=\!0.3$ is that the deviation of the cylinder axis from the $z$-axis for $\kappa\!=\!0.2$ is relatively larger  than that for $\kappa\!=\!0.3$.

To see the ordering of the variable $\sigma$ in more detail, we calculate the eigenvalues $\Sigma$ of the tensor order parameter defined by 
\begin{eqnarray}
\label{tensor-order} 
Q_{\mu\nu}=3\left(\langle\sigma_\mu\sigma_\nu\rangle-\delta_{\mu\nu}/3\right). 
\end{eqnarray}
If $\sigma$ is completely ordered, the largest eigenvalue $\Sigma_1$ becomes $\Sigma_1\!\to\! 1$, and the other two eigenvalues $\Sigma_2$ and $\Sigma_3$ are expected to be $\Sigma_{2,3}\!\to\! -0.5$.  We find from the data in Figs. \ref{fig-5} (c),(d) that the variable $\sigma$ is almost completely ordered, although the axis of $\sigma$ changes with increasing $L$. The completely ordered alignment of $\sigma$ in Fig. \ref{fig-5} (d) is expected from the fact that the assumed coefficient  $\lambda\!=\!3.33$ is relatively large.

\begin{figure}[h]
\centering
\includegraphics[width=11.5cm]{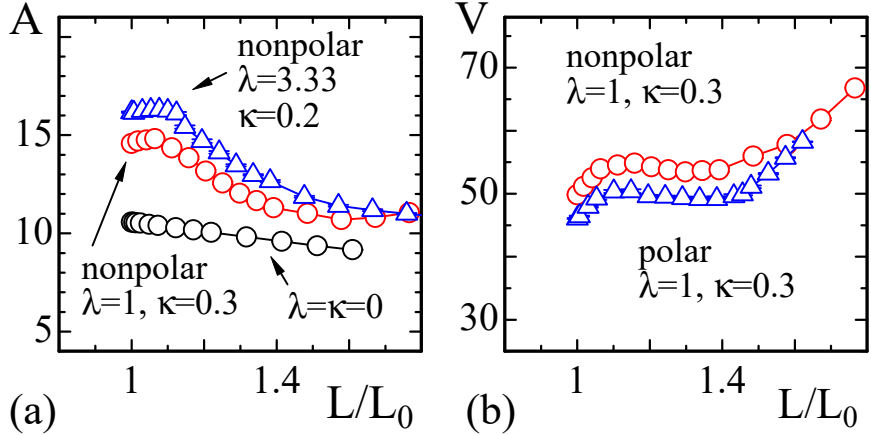}  

\caption{ (a) the mean sectional area
  $A(=\!V/L)$ vs. $L/L_0$, and (b) the volume $V$ vs.  $L/L_0$ for the
  same $\lambda$ and $\kappa$ as in Fig. \ref{fig-4}.
\label{fig-6} }
\end{figure}
Figure \ref{fig-6}(a) shows that the mean sectional area $A\!=\!V/L$
decreases as $L/L_0$ increases. The plateau of the volume $V$ in
\ref{fig-6}(b), except for the large $L/L_0$ region, reflects a
structural change.  Indeed, $V$ at $L/L_0\!\simeq\!1$ slightly
increases as $L/L_0$ increases, whereas $V$ remains constant in the
region of $L/L_0$ where $\tau$ has the plateau. More precisely, we
should recall that the tetrahedron resists deforming to be oblong
unless $\kappa\!\to\! 0$. For this reason, the tetrahedron shape (not
the size) is expected to remain almost unchanged when the size is
enlarged in the height direction, and consequently, the tetrahedron
volume increases with increasing height.  Thus, the volume $V$ of the
cylinder is expected to increase with increasing $L/L_0$ for
$\kappa\!\not=\!0$.  However, $V$ does not change in this manner and
remains almost constant. This constant $V$ represents a certain
lattice structural change that corresponds to the change of the variable $\sigma$ in our model.
 For the region $L/L_0\!>\!1.4$ in Fig. \ref{fig-6}(b), $V$ linearly increases with $L/L_0$. This implies that the bond length (and hence $S_1$)  also increases, and this is the reason why $\tau$ defined by Eq. (\ref{stress}) increases (Fig. \ref{fig-4}) in this region. We also note that the volume $V$ does not reflect the continuous transition between the nematic and isotropic phases.

\subsection{Elongation}\label{elong-simu}
As in the previous section, the tension coefficient is fixed to
$\gamma\!=\!1$. The symbol $L$ denotes the maximal diameter of the
oblong sphere in this section (whereas in the previous section, $L$
denotes the height of the cylinder). For the simulations of the elongation, we use spheres of size $(N,N_2)\!=\!(853,314)$, $(N,N_2)\!=\!(2423,1002)$, and $(N,N_2)\!=\!(4601,1402)$, where $N$ is the total number of vertices, $N_2$ is the total number of vertices on the surface, and $N$ includes $N_2$. The partition function for the spherical body is 
\begin{eqnarray}
\label{part-func-sphe} 
Z_{\rm sph}=\sum_\sigma \int^\prime \prod_{i=1}^N d{\bf r}_i \exp[-S({\bf r},\sigma)], 
\end{eqnarray}
where $\int^\prime$ denotes that the center of mass of the sphere is
fixed to the origin of ${\bf R}^3$. In sharp contrast to $Z_{\rm cyl}$
in Eq. (\ref{part-func-cyl}), $Z_{\rm sph}$ in
Eq. (\ref{part-func-sphe}) has no two-dimensional integrations. This
implies that no constraint except $\int^\prime$ is imposed on the
sphere.  As mentioned in Section \ref{model}, the self-avoiding
potential $U_{2D}$ is assumed in this case for the surface
triangles. Since this self-avoiding interaction in $U_{2D}$ is
non-local, the elongation simulation is relatively time consuming
compared to the soft-elasticity simulations in the previous
subsection. In Ref.  \cite{IC-Msquare-2015}, the two-dimensional
bending energy $S_2\!=\!\sum_{ij}(1\!-\!{\bf n}_i\cdot{\bf n}_i)$,
which is defined only for the surface, is examined rather than $S_2$ in Eq. (\ref{Disc-Eneg}), and the elongation phenomena are also observed. This two-dimensional $S_2$ is not included in the Hamiltonian for the simulations in this paper.

We note that the spherical body does not shrink to a point-like collapsed ball because of the scale invariance of the partition function as discussed in Section \ref{soft-simu}. Indeed, from the equation $\partial \log Z_{\rm sph}(\alpha)/\partial \alpha |_{\alpha=1}\!=\!0$ corresponding to Eq.(\ref{scale-inv-of-Z}), we obtain  $2\gamma \langle S_1\rangle\!-\!3(N\!-\!1)=0$, where  $\gamma\!=\!1$.  Thus, we have
\begin{eqnarray}
\label{scale-inv-sph} 
\langle S_1\rangle/N=3/2
\end{eqnarray}
in the limit of $N\!\to\! \infty$. Since $S_1$ is given by $S_1\!=\!(1/{4\bar N})\sum_{ij}\Gamma_{ij}\ell_{ij}^2$ and $\Gamma_{ij}\!=\!\sum_{\rm tet}\gamma_{ij}({\rm tet})$ is finite, the mean value of $\ell_{ij}^2$ becomes finite. This finiteness of $\ell_{ij}^2$ implies that the spherical body does not shrink to a small ball even without boundary conditions such as the upper and lower plates in the soft elasticity simulation in Section \ref{soft-simu}. 

\begin{figure}[h]
\centering
\includegraphics[width=11.5cm]{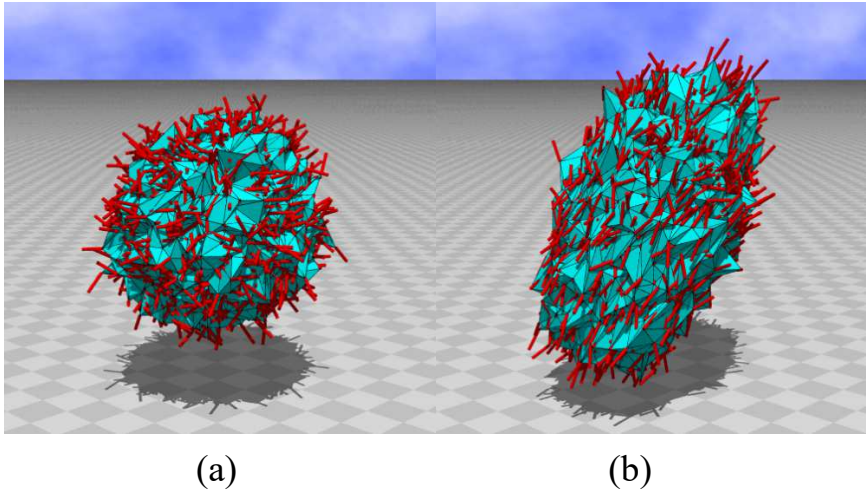}  
\caption{Snapshots of the elongation simulations for the non-polar model for (a) $\lambda\!=\!0.2$ (symmetric phase) and (b) $\lambda\!=\!0.427$ (elongated phase). These are obtained at $\kappa\!=\!0.1$, and the total number of vertices is $N\!=\!4601$. Small (red) cylinders represents the variable $\sigma$ defined at the vertices.
\label{fig-7}
 } 
\end{figure}
First, we present snapshots of symmetric and non-symmetric (elongated) spheres for the non-polar model in Figs. \ref{fig-7}(a),(b), where $\kappa\!=\!0.1$.  The variable $\sigma$ defined at the vertices is almost random for the symmetric phase in Fig. \ref{fig-7}(a), whereas it is almost aligned or ordered along the oblong direction for the elongated phase in Fig. \ref{fig-7}(b). There is no difference in the outside views between the polar and  non-polar models. 

There are several reasons for why the surface appears quite rough. One of the reasons is that the variable $\sigma$ is prohibited from being vertical to the edges of tetrahedron because of the divergence of $\gamma_{ij}$ in Eq. (\ref{surface-tension-coeff}). This divergence of  $\gamma_{ij}$ can be avoided on rough surfaces. Another reason is that the total number $N$ of vertices is not so large. If $N$ is large enough, then the surface appears relatively smooth.  It is also possible to smother the surface by including the two-dimensional bending energy on the surface into the Hamiltonian \cite{IC-Msquare-2015}. However, we should emphasize that the surface roughness such as the ones in the snapshots does not make any influences on the elongation phenomenons in the model. The important point is that the tetrahedron does not always represent the actual structure of LCE, it is introduced only for the discretization of spherical body such that its size is negligible compared with the size of spherical body in the limit of $N\to\infty$.

\begin{figure}[h]
\centering
\includegraphics[width=11.5cm]{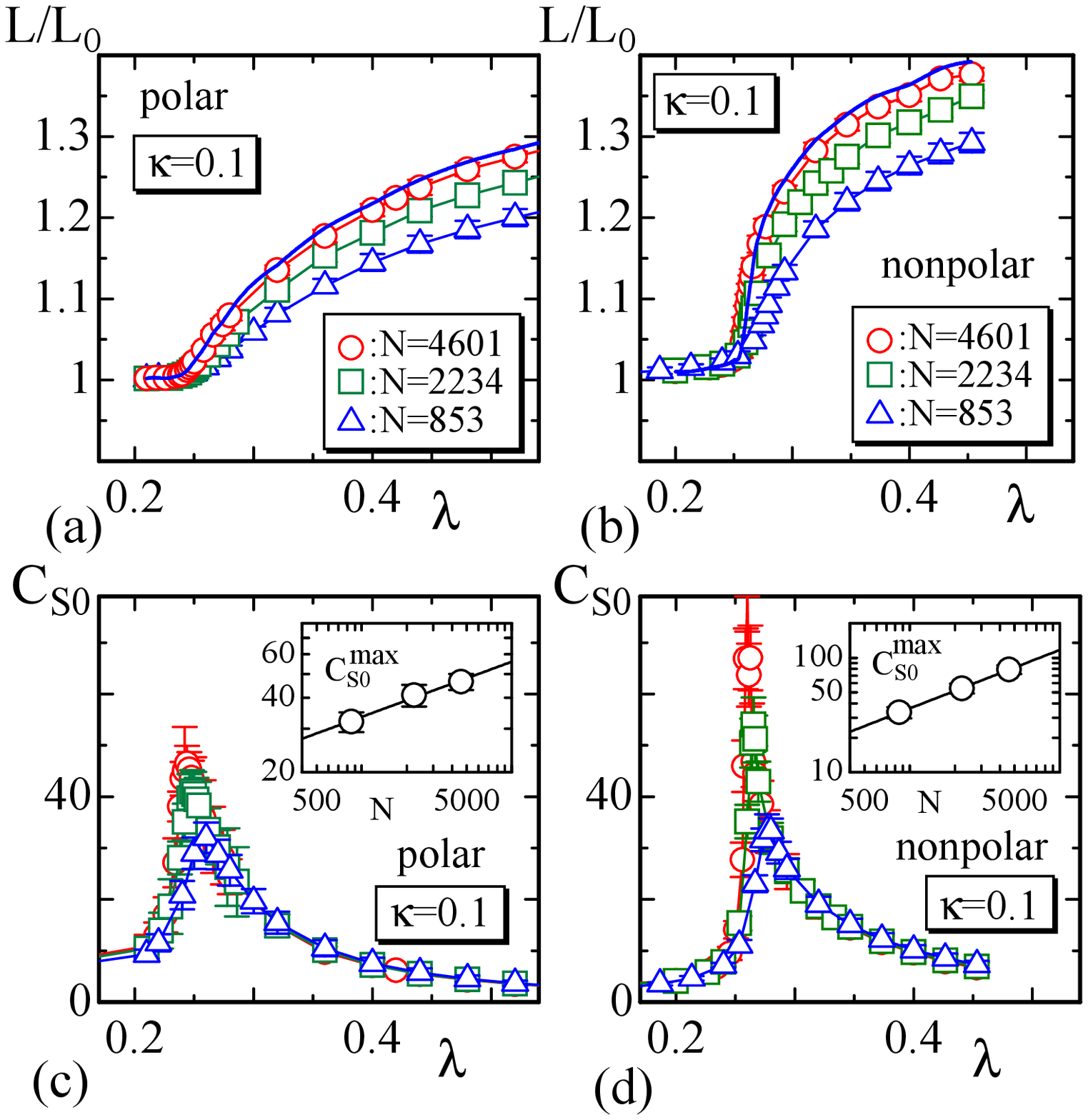}  
\caption{$L/L_0$ vs. $\lambda$ of the (a) polar and (b) non-polar models, and the variance $C_{S_0}$ vs. $\lambda$ of the (c) polar and (d) non-polar models. $\kappa\!=\!0.1$. The thick solid lines in (a) and (b) are those for $N\!\to\!\infty$ obtained from the linear fitting of data by $1/N$.
\label{fig-8}
 } 
\end{figure}
We plot the strain $L/L_0$ vs. $\lambda$ in Figs. \ref{fig-8}(a),(b), where $L_0$ is $L$ for $\lambda\!=\!0$. For $\lambda\!=\!0$, the sphere is not elongated, and hence, it remains symmetric. 
The stiffness is fixed to $\kappa\!=\!0.1$ in both the polar and non-polar cases. Note that large (small) $\lambda$ corresponds to low (high) temperature since $\lambda$ has units of $k_BT$.  Therefore, the increasing $\lambda$ along the horizontal axis from the origin to the right direction simply corresponds to the decreasing temperature. The solid lines connecting the data symbols are obtained by interpolating the data for the lattices of size $N\!=\!853$, $N\!=\!2234$, and $N\!=\!4601$ with the Legendre polynomial. By fitting these interpolated data $L/L_0$ linearly against $1/N$, we obtain the thick solid lines corresponding to the data in the limit of $N\!\to\!\infty$.  Note that $L/L_0$ in the large $\lambda$ region becomes relatively smaller for larger $\kappa$, such as $\kappa\!=\!0.3$ and $\kappa\!=\!0.5$, which are not plotted. The large $\kappa$ protects  the sphere from elongating. 
The results of $L/L_0$ vs. $\lambda$ plotted in Fig. \ref{fig-8} are consistent with  those of the experimental ones reported in \cite{V-Domenici-2012,Greve-MacromCP-2001} (see Fig. \ref{fig-1}(b)), although our results are obtained under zero external tensile stress, where the elongation axis is spontaneously determined.  In the experiments reported in \cite{V-Domenici-2012,Greve-MacromCP-2001},  the curve of $L/L_0$ vs. the temperature $T$ is measured, whereas in Fig. \ref{fig-8}, the curve of $L/L_0$ vs. $\lambda$ is plotted. This is the major difference between the data in Fig. \ref{fig-8} and  the experimental data.

The elongation is caused by the transition between the disordered (=  spherical) and ordered (= elongated)  phases of $\sigma$. This transition can be reflected in the variance defined by 
\begin{eqnarray} 
\label{specific-h}
C_{S_0}=(1/N)\left(\langle S_0^2\rangle-\langle S_0\rangle^2 \right).
\end{eqnarray} 
This variance $C_{S_0}$ can be called specific heat corresponding to the energy $S_0$, because
of the relations  $-(\partial /\partial\lambda) \log Z\!=\!\langle S_0\rangle$ and $-(\partial^2 /\partial\lambda^2) \log Z\!=\!\langle S_0^2\rangle\!-\!\langle S_0\rangle^2$. 
  Note that the coefficient $\lambda^2$ is not included in $C_{S_0}$ of Eq. (\ref{specific-h}), 
however the critical behavior of $C_{S_0}$ is independent of this coefficient. 
Although $S_0/N$ (which is not plotted) varies almost
continuously,  we observe
from Figs. \ref{fig-8}(c),(d) that $C_{S_0}$ has a peak $C_{S_0}^{\rm
  max}$, which increases with increasing $N$. Note that the critical
point $\lambda_c(N)$, where $C_{S_0}$ has the peak $C_{S_0}^{\rm max}$,
is the point where $L/L_0$ suddenly increases from  $L/L_0\!=\!1$ in
both the polar and non-polar cases. 
We obtain the scaling coefficient $\alpha$ in $C_{S_0}^{\rm max}\sim N^\alpha$ such that
\begin{eqnarray} 
\label{scaling-coeff-1}
&&\alpha=0.22\pm 0.07 \quad ({\rm polar}), \nonumber \\ 
&&\alpha=0.51\pm 0.08 \quad ({\rm nonpolar}), 
\end{eqnarray} 
for $\kappa\!=\!0.1$. This implies that the model has a continuous
transition for both the polar and non-polar cases.  These results confirm the continuous nature of the transition between the nematic and isotropic phases in LCE \cite{V-Domenici-2012}. Therefore, we consider that the FG model in this paper corresponds to  the main chain LCE, because  the main (side) chain LCE undergoes a continuous (discontinuous) transition in experiments \cite{V-Domenici-2012}. 
Although the order of the transition remains continuous or second order in both models, the coefficient $\alpha$ of the non-polar model is larger  than  that of the polar model. This is consistent to the fact that the change of $L/L_0$ at the transition point is more abrupt in the non-polar model than in the polar model as we have confirmed in Figs. \ref{fig-8}(a), (b).

\begin{figure}[h]
\centering
\includegraphics[width=11.5cm]{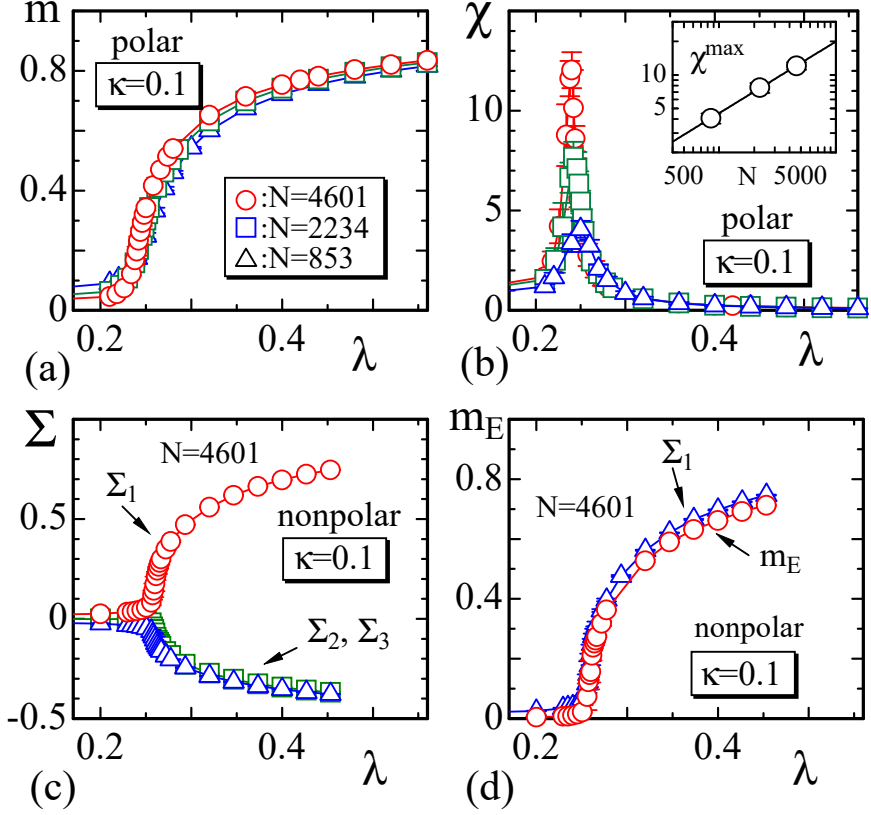}  
\caption{(a) $m$ vs. $\lambda$ and (b) the susceptibility $\chi$ vs. $\lambda$ of the polar model. The peak of susceptibility $\chi^{\rm max}$  vs. $N$ is plotted in a log-log scale in the small window in (b).  (c) The eigenvalues $\Sigma_{1,2,3}$ of the tensor order parameter $Q_{\mu\nu}$ in Eq. (\ref{tensor-order}) vs. $\lambda$, and (d) the parameter $m_E$ (and $\Sigma_1$) vs.  $\lambda$ of the non-polar model. 
\label{fig-9}
}
\end{figure}
The order parameter $m$ of the transition for the polar model is given by
\begin{eqnarray} 
\label{order-param}
m=    \langle\sigma\rangle  \quad ({\rm polar}). 
\end{eqnarray} 
This magnetization $m$ changes just like in the ferro-magnetic transitions (see Fig. \ref{fig-9}(a)). The susceptibility  
\begin{eqnarray} 
\chi=N\left(\langle \sigma^2\rangle-\langle \sigma\rangle^2 \right)
\end{eqnarray} 
 has a peak at the transition point, where $L/L_0$ begins to
 increase. The peak value $\chi^{\rm max}(N)$ is expected to scale
 according to $\chi^{\rm max}(N)\!\sim\! N^\nu$; indeed, we have
 $\nu\!=\!0.64\!\pm\!0.08$ (Fig. \ref{fig-9}(b)). This implies that
 the variable $\sigma$ also plays an important role in the elongation phenomenon of the model.

To see the ordering of the variables $\sigma$ for the non-polar model, we calculate 
 the eigenvalues of the tensor order parameter in Eq. (\ref{tensor-order}). The eigenvalues $\Sigma_{1,2,3}(\Sigma_1\!>\!\Sigma_2\!>\!\Sigma_3)$ abruptly change at the transition point (Fig. \ref{fig-9}(c)). We also find that $\Sigma_{1,2,3}\!\to\!0$ for $\lambda\!\to\!0$ and $\Sigma_1\!\to\!1$, $\Sigma_{2,3}\!\to\!-0.5$ for $\lambda\!\to\!\infty$. This implies that $\sigma$ becomes random (ordered) in the limit of  $\lambda\!\to\!0$ ($\lambda\!\to\!\infty$) at the transition point $\lambda_c(\simeq\! 0.25)$.

To evaluate the ordering of the variables $\sigma$ along the elongation axis, we calculate the parameter
\begin{eqnarray} 
m_E=(1/2)\left(3\langle(\sigma\cdot{\bf t}_E)^2\rangle-1\right) \quad({\rm nonpolar})
\end{eqnarray} 
for the non-polar model. In this expression, the elongation axis ${\bf t}_E$ is evaluated by 
\begin{eqnarray} 
&&{\bf t}_E=\frac{{\bf r}_I-{\bf r}_J}{|{\bf r}_I-{\bf r}_J|}, \nonumber\\
&& |{\bf r}_I-{\bf r}_J|={\rm Max}\{|{\bf r}_i-{\bf r}_{j}|(i,j = {\rm vertices\; on \; the \; surface})\},
\end{eqnarray} 
where $I$ and $J$ denote the vertices on the surface such that the distance $|{\bf r}_I-{\bf r}_J|$ is the maximum. If this axis ${\bf t}_E$ is identical to the ordered axis of $\sigma$, $m_E$ is identified to $\Sigma_1$. We find from  Fig. \ref{fig-9}(d) that the variable $\sigma$ almost aligns along the elongated direction as we have confirmed in the snapshots in Fig. \ref{fig-7}(b).

\begin{figure}[h]
\centering
\includegraphics[width=11.5cm]{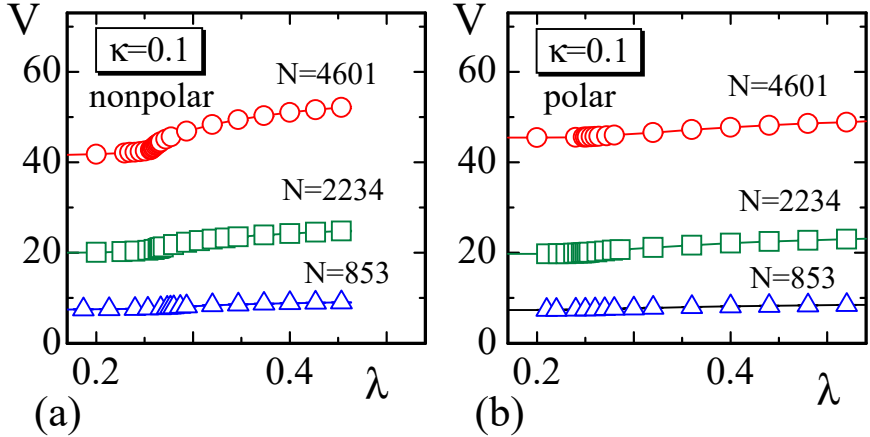}  
\caption{The volume $V$ vs. $\lambda$ for the (a) non-polar and (b) polar models.
\label{fig-10} } 
\end{figure}

The continuous transition of $\sigma$ is reflected in the volume $V$. Indeed, $V$ continuously changes at the transition point  in both the polar and non-polar models. However, the change of $V$ is relatively small even at the transition point (Figs. \ref{fig-10}(a),(b)), and the continuous transition  is not reflected in the variance  $C_V$ of $V$ in both models (these are not plotted).

Next, a constraint on the volume $V$, such as 
\begin{eqnarray} 
\label{volume_constraint}
V=V_0={\rm constant},
\end{eqnarray} 
is imposed on the simulations. In the simulations for the elongation
in  Figs. \ref{fig-8} and \ref{fig-9}  (as well as in Fig. \ref{fig-7}), no constraint, including the one in Eq. (\ref{volume_constraint}), is imposed.  
The reason why this constraint is imposed here is to determine whether the elongation can be observed without the volume change. In fact, we have observed that the elongation accompanies a small variation of $V$ as in Fig. \ref{fig-5}(d) and in Figs.  \ref{fig-10}(a),(b). In the simulations  without the constraints, the volume changes in the MC process  only when the vertices on the surface are updated, whereas it remains unchanged in the update of vertices inside the body. Therefore, if the volume is rigorously fixed under the constraint of Eq. (\ref{volume_constraint}), the vertices on the surface can move only in specific directions such that the surface shape remains unchanged from the initial smooth cylinder. For this reason, we impose a constraint  on the volume $V$ such that
\begin{eqnarray} 
\label{volume-constant}
V_0- {\it \Delta}V\leq V\leq V_0+{\it \Delta}V
\end{eqnarray} 
where ${\it \Delta}V$ is the volume of the regular tetrahedron, the bond length of which is given by the mean bond length in the equilibrium configuration. The equilibrium bond length is expected to remain constant from the relation $S_1/N\!=\!3/2$, which comes from the scale invariance of $Z_{\rm sph}$. From this, we find that ${\it \Delta}V/V_0$ becomes very small, and the rate of acceptance for the vertex moving on the surface is almost uninfluenced by this constraint. Hence, the constraint of Eq. (\ref{volume-constant}) is accurate and meaningful, and this technique is the same as the one used in the enclosed-volume constant simulations for membranes in \cite{Koibuchi-etal-JOMC2016}.

\begin{figure}[h]
\centering
\includegraphics[width=11.5cm]{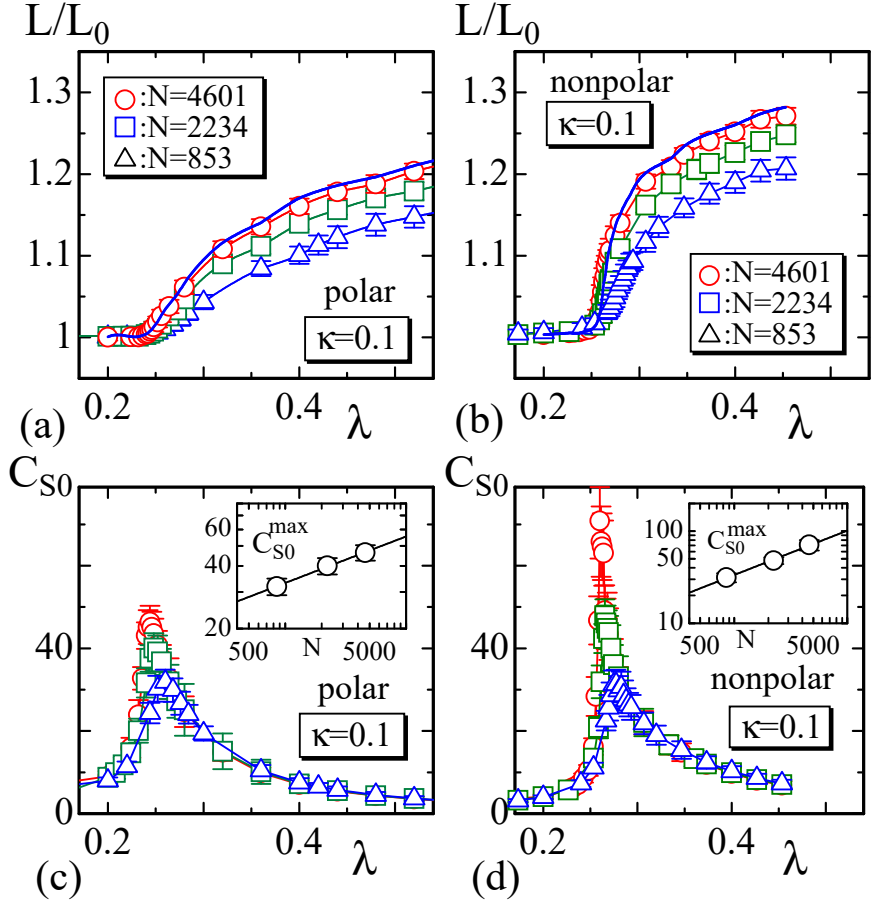}  
\caption{The volume constant simulation results: $L/L_0$ vs. $\lambda$ of the (a) polar and (b) on-polar models, and $C_{S_0}$  vs. $\lambda$ of the (c) polar and (d) non-polar models. The thick solid lines in (a) and (b) correspond to $L/L_0$ vs. $\lambda$ for $N\!\to\!\infty$.  
\label{fig-11}
} 
\end{figure}
The results of $L/L_0$ vs. $\lambda$ in both the polar and non-polar cases for $\kappa\!=\!0.1$ in Figs. \ref{fig-11}(a)--(d) are almost identical to those in Figs. \ref{fig-8}(a)--(d) obtained from the model without the constraint on $V$. The phase transition between the symmetric and elongated phases also remains unchanged. We have $C_{S_0}^{\rm max}(N)\!\sim\! N^\alpha$, $\alpha\!=\!{0.48\!\pm\!0.10}$ for the case of non-polar and $\kappa\!=\!0.1$ (Fig. \ref{fig-11}(d)). This component $\alpha$ is identified to the second of Eq. (\ref{scaling-coeff-1}) within the error.

\section{Summary and conclusion}\label{Discussion}
In this paper,  we introduce a new model for a $3D$ liquid crystal elastomer (LCE). This model is constructed on the basis of Finsler geometry (FG). Regarding the soft elasticity and elongation phenomena of LCE, we confirm that the Monte Carlo data are consistent with the existing experimental results. From these numerical simulations, we find that the mechanism for the anisotropy in the FG model is deeply connected with the interaction of $\sigma$ with the position ${\bf r}$ of  $\sigma$. This interaction is coarse grained and implemented in the model via the Finsler metric.

We provide speculative comments on several possible applications of FG modeling. 

The first is the
deformation of a thin LCE \cite{Camacho-etal-Nature2004}. The
mechanism of a deformation of a thin LCE under non-uniform
illumination of visible light can be understood in the framework of FG
modeling. For such thin LCE, the temperature dependence of physical
quantities can be evaluated with the parameter $\lambda$ in our model. The second is the deformation of LCE under external electric fields \cite{Urayama-PRE2005}. The variables $\sigma$ is aligned by an electric field along or vertical to the electric field, and deformation of LCE is expected to be independent of how $\sigma$ is aligned.  Finally, the so-called J-shaped stress-strain diagram of biological materials, such as blood vessels
and skin, can also be considered in the scope of FG modeling \cite{Greven-JOM-1995}.




\noindent
{\bf Acknowledgment}

The author H.K. acknowledges Giancarlo Jug for comments and Andrei Makisimov for discussions during IC-Msquare 2015, and he also acknowledges Andrey Shobukhov for comments. The authors acknowledge Satoshi Usui and Eisuke Toyoda for computer analyses. This work is supported in part by JSPS KAKENNHI Number 26390138.

\appendix
\section{$3D$ Finsler metric}\label{FG_model}
In this Appendix, we show how to obtain the $3D$ Finsler metric in Eq. (\ref{Finsler_metric})
  with the help of FG modeling in a
self-contained manner.  The
 anisotropy in LCE, such as oblong shape for example, is considered
to be connected with the internal molecular structure, such as the
direction of  liquid crystal molecules. This intuitive
picture for the anisotropy is understood in the context of the Finsler geometry model.

 The Hamiltonian of the LCE model includes the Gaussian energy $S_1$ described in Eq. (\ref{cont_S}). 
 In this $S_1$, ${\bf r}(\in {\bf R}^3)$ denotes the LCE position,   $g^{ab}$ is the inverse of the metric $g_{ab}$, which is a $3\times 3$ matrix,  and $g$ is its determinant.   

\begin{figure}[h]
\centering
\includegraphics[width=11.5cm]{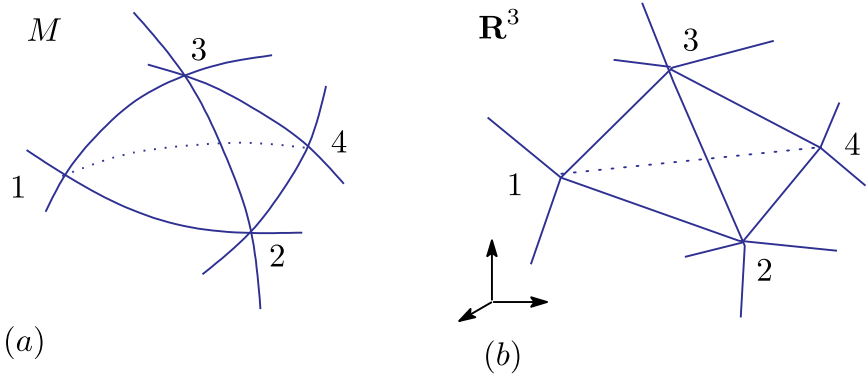}   
\caption{ (a) A smooth  tetrahedron $1234$ in $M$, and (b) the piece-wise linear  tetrahedron $1234$ in ${\bf R}^3$.
 \label{fig-A1}} 
\end{figure}
Since the LCEs are considered to be a $3D$ body in ${\bf
  R}^3$, the induced metric $g_{ab}\!=\!\partial_a {\bf r}\!\cdot\!
\partial_b {\bf r}$ can be assumed for the LCE model
\cite{FDavid-SMS2004}. Here, however, we slightly generalize the
metric function to implement the Finsler metric. For this purpose, it
is convenient to consider that ${\bf r}$ is a mapping from a
three-dimensional parameter space $M$ to ${\bf R}^3$ such that ${\bf
  r}:M\ni (x_1,x_2,x_3)\mapsto {\bf r}(x_1,x_2,x_3)\in{\bf R}^3$  (see
Figs. \ref{fig-A1}(a),(b)). This space $M$ is considered as a three-dimensional manifold, which is locally identified with a domain in ${\bf R}^3$. The elements of $g_{ab}$ are functions on $M$, and $g_{ab}$ is assumed to be positive definite ($\Leftrightarrow \sum_{ab}g_{ab}v_av_b\!>\!0$ for all $(v_1,v_2)\!\not=\!(0,0)$).

$M$ is called a Finsler space if $M$ is equipped with a Finsler function $L$ \cite{Matsumoto-SKB1975,Bao-Chern-Shen-GTM200}. Let $C$ be a curve on $M$ such that $ C\ni t\mapsto x(t)\in M$; then, the Finsler length $s$ along $C$ is defined using $L$ such that   
\begin{equation}
\label{F_length}
s=\int_{t_0}^t L(x,y)dt\quad (\Leftrightarrow \frac {ds} {dt} =L(x,y)),
\end{equation}
where $x\!=\!(x_1,x_2,x_3)$ and $y\!=\!(y_1,y_2,y_3)\!=\!(dx_1/dt,dx_2/dt,dx_3/dt)$
denote a point on $C$ and a tangential vector along $C$, respectively. The original Finsler function $L$ is given by
\begin{equation}
\label{example}
L(x(t),y(t))=\sqrt{\sum_i y_i^2}/\vert{\bf v}\vert,
\end{equation}
where ${\bf v}$ and $|{\bf v}|=\sqrt{\sum_i(dx_i/ds)^2}$ are a vector
along $C$ and its length with respect to the special parameter $s$,
respectively \cite{Matsumoto-SKB1975}.  This $L$ satisfies
$ds/dt\!=\!L$ (see Ref. \cite{Koibuchi-Sekino-PhysicaA2014} in more
detail on this point). Note that ${\bf v}$ is a tangential vector of
$C$ with respect to the Finsler length $s$ along $C$, and hence, the
length $|{\bf v}|$ plays a role of unit Finsler length. Since the
integral $\int \sqrt{\sum_i y_i^2}dt$ provides the ordinary length of
$C$, $L$ is considered to be the ratio of the ordinary length unit and
the Finsler length unit along $C$. This ratio depends on the direction
of $C$ (or ${\bf v}$) if $|{\bf v}|$ depends on the direction.

The problem is where  $|{\bf v}|$ comes from.  One answer to this problem is that 
\begin{equation}
\label{vector_on_M}
{\bf v}=(\sigma \cdot {\bf t}) {\bf t},
\end{equation}
where $\sigma(\in S^2)$ is a three-dimensional unit vector
corresponding to a liquid crystal (LC) molecule and ${\bf t}$ is a unit tangential vector along the tetrahedron edge. For the non-polar interaction, $-\sigma$ and $\sigma$ are identified. It is natural to consider that $\sigma$ is given at the vertices of the tetrahedrons in ${\bf R}^3$ because the tetrahedrons correspond to an LCE. This implies that ${\bf v}\in {\bf R}^3$, although it should originally belong to $M$. The important point to note is that we simply assume the same length ${|\bf v}|$ on the smooth tetrahedron in $M$ to define the Finsler function $L$.  Note that ${\bf v}\!=\!{\bf 0}$ for any $\sigma$ with $\sigma \cdot {\bf t}\!=\!0$, and in this case, the Finsler function $L$ is not defined along the direction corresponding to this ${\bf t}$.

Let $v_{12}$ be the component of ${\bf v}$ along the edge (or bond) $12$ of the tetrahedron, as shown in Fig. \ref{fig-A1}(b). Thus, we have 
\begin{eqnarray}
\label{vector_on_bonds}
&&v_{12}=|{\bf v}\cdot {\bf t}_{12}|=|\sigma_1\cdot {\bf t}_{12}|, \quad v_{13}=|{\bf v}\cdot {\bf t}_{13}|=|\sigma_1\cdot {\bf t}_{13}|, \nonumber \\
&&v_{14}=|{\bf v}\cdot {\bf t}_{14}|=|\sigma_1\cdot {\bf t}_{14}|,
\end{eqnarray}
where $\sigma_i$ is $\sigma$ at vertex $i$ and ${\bf t}_{ij}$ is the
unit tangential vector along the bond $ij$ at vertex $i$. Note that
$v_{ij}\!\not=\!v_{ji}$ in general. Using these $v_{ij}$, we calculate
$L$ on the smooth tetrahedron 1234 in Fig. \ref{fig-A1}(a) as
follows. First, we assume that the local coordinate origin of the
smooth tetrahedron is at vertex 1, and then bonds $12$,$13$ and $14$
correspond to the coordinate axes $x_1$,$x_2$ and $x_3$. Therefore, the
discrete Finsler function $L_{12}$ on the $x_1$ axis of the smooth
tetrahedron is given by $L_{12}\!=\!\int(dx_1/dt)dt/v_{12}\!=\!\int
dx_1/v_{12}\!=\!1/v_{12}$, where $\int dx_1\!=\!1$ is assumed for bond
$12$. We also have $L_{13}\!=\!1/v_{13}$, $L_{14}\!=\!1/v_{14}$ for bonds $13$ and $14$. We assume that
$g_{ab}$ is the Euclidean metric $\delta_{ab}$ on the smooth tetrahedron
in Fig. \ref{fig-A1}(a), in which the coordinate origin is at vertex $1$; then, by replacing the elements of $\delta_{ab}$ with the Finsler length squares $L_{12}^2$, $L_{13}^2$  and $L_{14}^2$, we have the Finsler metric in Eq. (\ref{Finsler_metric}).

\section{Physical unit of stress $\tau$ and Young modulus $E$}\label{formula-for-tau}
We present the expression for experimental Young modulus $E$ obtained by using the simulation results $\tau(L)$ and experimental data $\tau_{\rm exp}$. For this purpose, it is convenient to introduce the notion of lattice spacing $a[m]$, which is fixed to $a\!=\!1$ under the unit of $k_BT\!=\!1$ in the simulations. All physical quantities that have the unit of length should be multiplied by $a$.  The inverse temperature $\beta\!=\!1/k_BT$ is also fixed to $\beta\!=\!1 (\Leftrightarrow k_BT\!=\!1)$ in the simulations, and as a consequence $\gamma S_1$ has the unit of $k_BT$ in Eq. (\ref{stress}) for example. 
 By including these parameters $a$, $\beta$ and $k_BT$ in the calculation of scale invariant property of the model introduced in Subsection \ref{soft-simu}, we have
\begin{eqnarray}
\label{experim-stress} 
\tau_{\rm exp}(E)&=&\sqrt{\frac{\left(2\gamma\langle S_1\rangle-2N_{2d}-3N_{3d}\right) k_BT E}{{LA_0a^3}}} \nonumber \\
&=&\sqrt{\frac{2\gamma\langle S_1\rangle-2N_{2d}-3N_{3d}}{{LA_0}}}\sqrt{\frac{k_BTE}{{a^3}}}\nonumber \\
&\simeq&\sqrt{\frac{2\gamma\langle S_1\rangle-2N_{2d}-3N_{3d}}{{LA_0}}}\sqrt{\frac{4\times 10^{-21}E}{{a^3}}}\nonumber \\
&=&\sqrt{\frac{2\gamma\langle S_1\rangle-2N_{2d}-3N_{3d}}{{LA_0}}}\sqrt{E}.
\end{eqnarray}
In the second line of Eq. (\ref{experim-stress}), $k_BT$ is replaced by $k_BT\!\simeq 4\times 10^{-21}[Nm]$, where the room temperature is assumed for $T$.  It should be noted that $a$ is given by $4\!\times\!10^{-21}\!=\!a^3 (\Leftrightarrow a\!\simeq\!1\!\times\! 10^{-7}[m])$ if the unit of $k_BT$ is explicitly used.  This $\tau_{\rm exp}(E)$ has the unit ${\rm [Pa]}$ if $E$ is given with the unit of ${\rm [Pa]}$.  Using the fact that the simulation result $\tau(L)$ is obtained from this $\tau_{\rm exp}(E)$ in Eq. (\ref{experim-stress}) by replacing $E$ with $E_{\rm sim}(=\!1{\rm [Pa]})$, we have 
\begin{eqnarray}
\label{simulation-stress} 
\tau(L)=\tau_{\rm exp}(E_{\rm sim})=\sqrt{\frac{2\gamma\langle S_1\rangle-2N_{2d}-3N_{3d}}{{LA_0}}}\sqrt{E_{\rm sim}}.
\end{eqnarray}
Note that the right hand side (= the final expression)  is exactly identical to $\tau(L)$ in Eq. (\ref{stress}) if $E_{\rm sim}$ is given by $E_{\rm sim}\!=\!1$. From Eqs. (\ref{experim-stress}) and (\ref{simulation-stress}),  we obtain $\tau(L)/\sqrt{E_{\rm sim}}\!=\!\tau_{\rm exp}/\sqrt{E}$. Thus, we finally have
\begin{eqnarray}
\label{Young-modulus}
E=\left(\tau_{\rm exp}/\tau(L)\right)^2 E_{\rm sim} =\left(\tau_{\rm exp}/\tau(L)\right)^2\; {\rm [Pa]},
\end{eqnarray}
where $E_{\rm sim}\!=\!1{\rm [Pa]}$ is assumed. 
This is a relation between the simulation data $\{\tau(L),E_{\rm sim}(=\!1)\}$ and experimental data  $\{\tau_{\rm exp},E\}$, where the unit of these quantities is ${\rm [Pa]}$.


\end{document}